\documentstyle[psfig]{mn}

\newif\ifAMStwofonts

\input{epsf}

\ifoldfss
  \ifCUPmtlplainloaded \else
    \NewTextAlphabet{textbfit} {cmbxti10} {}
    \NewTextAlphabet{textbfss} {cmssbx10} {}
    \NewMathAlphabet{mathbfit} {cmbxti10} {} 
    \NewMathAlphabet{mathbfss} {cmssbx10} {} 
  \fi
  \ifAMStwofonts
    \ifCUPmtlplainloaded \else
      \NewSymbolFont{upmath} {eurm10}
      \NewSymbolFont{AMSa} {msam10}
      \NewMathSymbol{\upi}     {0}{upmath}{19}
      \NewMathSymbol{\umu}     {0}{upmath}{16}
      \NewMathSymbol{\upartial}{0}{upmath}{40}
      \NewMathSymbol{\leqslant}{3}{AMSa}{36}
      \NewMathSymbol{\geqslant}{3}{AMSa}{3E}

      \let\leq=\leqslant 
      \let\geq=\geqslant 
    \fi
  \fi
\fi 

\ifnfssone
  \newmathalphabet{\mathit}
  \addtoversion{normal}{\mathit}{cmr}{m}{it}
  \addtoversion{bold}{\mathit}{cmr}{bx}{it}
  \newmathalphabet{\mathbfit} 
  \addtoversion{normal}{\mathbfit}{cmr}{bx}{it}
  \addtoversion{bold}{\mathbfit}{cmr}{bx}{it}
  \newmathalphabet{\mathbfss} 
  \addtoversion{normal}{\mathbfss}{cmss}{bx}{n}
  \addtoversion{bold}{\mathbfss}{cmss}{bx}{n}
  \ifAMStwofonts
    \ifCUPmtlplainloaded \else
      \UseAMStwoboldmath
      \makeatletter
      \new@mathgroup\upmath@group
      \define@mathgroup\mv@normal\upmath@group{eur}{m}{n}
      \define@mathgroup\mv@bold\upmath@group{eur}{b}{n}
      \edef\UPM{\hexnumber\upmath@group}
      \new@mathgroup\amsa@group
      \define@mathgroup\mv@normal\amsa@group{msa}{m}{n}
      \define@mathgroup\mv@bold\amsa@group{msa}{m}{n}
      \edef\AMSa{\hexnumber\amsa@group}
      \makeatother
      \mathchardef\upi="0\UPM19
      \mathchardef\umu="0\UPM16
      \mathchardef\upartial="0\UPM40
      \mathchardef\leqslant="3\AMSa36
      \mathchardef\geqslant="3\AMSa3E

      \let\leq=\leqslant 
      \let\geq=\geqslant 
    \fi
  \fi
\fi 

\ifnfsstwo
  \DeclareMathAlphabet{\mathbfit}{OT1}{cmr}{bx}{it}
  \SetMathAlphabet\mathbfit{bold}{OT1}{cmr}{bx}{it}
  \DeclareMathAlphabet{\mathbfss}{OT1}{cmss}{bx}{n}
  \SetMathAlphabet\mathbfss{bold}{OT1}{cmss}{bx}{n}
  \ifAMStwofonts
    \ifCUPmtlplainloaded \else
      \DeclareSymbolFont{UPM}{U}{eur}{m}{n}
      \SetSymbolFont{UPM}{bold}{U}{eur}{b}{n}
      \DeclareSymbolFont{AMSa}{U}{msa}{m}{n}
      \DeclareMathSymbol{\upi}{0}{UPM}{"19}
      \DeclareMathSymbol{\umu}{0}{UPM}{"16}
      \DeclareMathSymbol{\upartial}{0}{UPM}{"40}
      \DeclareMathSymbol{\leqslant}{3}{AMSa}{"36}
      \DeclareMathSymbol{\geqslant}{3}{AMSa}{"3E}

      \let\leq=\leqslant 
      \let\geq=\geqslant 
    \fi
  \fi
\fi 

\ifCUPmtlplainloaded \else
  \ifAMStwofonts \else 
    \def\upi{\pi}
    \def\umu{\mu}
    \def\upartial{\partial}
  \fi
\fi

\title{Properties of Galaxy Clusters: Mass and Correlation Functions  }
\author[F.Governato \etal ]
       {F. Governato$^{\bf 1}$,
		A. Babul$^{\bf 2,3}$,
		T. Quinn$^{\bf 3}$, 
		P. Tozzi$^{\bf 4,5}$,\\ \\
      {\LARGE	C.M. Baugh$^{\bf 1}$,
		N. Katz$^{\bf 6}$,
		G. Lake$^{\bf 3}$} \\ \\
        $^{\bf 1}$Physics Department, Science Labs, South Road, 
        Durham, DH1 3LE, UK, Fabio.Governato@durham.ac.uk \\ 
        $^{\bf 2}$ Dept. of Physics \& Astronomy, Elliot Building, Univ. of 
         Victoria, Victoria, BC, Canada, V8P 1A1 \\  
        $^{\bf 3}$ Astronomy Department, University of Washington, Seattle
         WA, USA   \\
        $^{\bf 4}$ Space Telescope Science Institute, 3700 San Martin Drive, 
         Baltimore, MD 21218, USA \\
        $^{\bf 5}$ II Universit\'a di Roma, Tor Vergata, Roma, Italia \\
        $^{\bf 6}$ Dept. of Physics \& Astronomy, 517 Lederle Graduate
        Research Tower, Univ. of Massachusetts, Amherst, MA 01003-4525, USA}

\date{submitted to MNRAS}

\pagerange{\pageref{firstpage}--\pageref{lastpage}}
\pubyear{1998}

\begin{document}

\maketitle

\label{firstpage}

\def\etal  {\it {et al.} \rm}

\begin{abstract}
We analyse parallel N-body simulations of three Cold Dark Matter (CDM)
universes to study the abundance and clustering of galaxy
clusters. The simulation boxes are $500 h^{-1} $Mpc on a side and
cover a volume comparable to that of the forthcoming Sloan Digital Sky
Survey. The use of a treecode algorithm and 47 million particles
allows us at the same time to achieve high mass and force resolution.
We are thus able to make robust measurements of cluster properties
with good number statistics up to a redshift larger than unity.  We
extract halos using two independent, public domain group finders
designed to identify virialised objects -- `Friends-of-Friends' (Davis
\etal 1985) and `HOP' (Eisenstein \& Hut 1998) -- and find consistent
results.  The correlation function of clusters as a function of mass
in the simulations is in very good agreement with a simple analytic
prescription based upon a Lagrangian biasing scheme developed by Mo \&
White (1996) and the Press-Schechter (PS) formalism for the mass
function.  The correlation length of clusters as a function of their
number density, the $R_{0}$--$D_{c}$ relation, is in good agreement
with the APM Cluster Survey in our open CDM model.  The critical
density CDM model (SCDM) shows much smaller correlation lengths than
are observed.  We also find that the correlation length does not grow
as rapidly with cluster separation in any of the simulations as
suggested by the analysis of very rich Abell clusters.  Our SCDM
simulation shows a robust deviation in the shape and evolution of the
mass function when compared with that predicted by the PS
formalism. Critical models with a low $\sigma_8$ normalization or
small shape parameter $\Gamma$ have an excess of massive clusters
compared with the PS prediction.  When cluster normalized, the SCDM
universe at z $= 1$ contains 10 times more clusters with temperatures
greater than $7$keV, compared with the Press \& Schechter
prediction. The agreement between the analytic and N-body mass
functions can be improved, for clusters hotter than 3 keV in the
critical density SCDM model, if the value of $\delta_{c}$ (the
extrapolated linear theory threshold for collapse) is revised to be $
\delta_{c}(z) = 1.685 \left[ (0.7/\sigma_{8}) (1+z) \right]^{-0.125} $
($\sigma_{8}$ is the {\it rms} density fluctuation in spheres of
radius $8 h^{-1}$Mpc). Our best estimate for the amplitude of
fluctuations inferred from the local cluster abundance for the SCDM
model is $\sigma_{8} = 0.5 \pm 0.04$. However, the discrepancy between
the temperature function predicted in a critical density universe and
that observed at $z=0.33$ (Henry \etal 1998) is reduced by a modest
amount using the modified Press-Schechter scheme. The discrepancy is
still large enough to rule out $\Omega_{0} = 1$, unless there are
significant differences in the relation between mass and temperature
for clusters at high and low redshift.
\end{abstract}


\section{Introduction}
Clusters of galaxies, by virtue of being both relatively rare objects
and the largest gravitationally bound systems in the Universe, provide
stringent constraints on theories of structure formation.  The two
cluster properties that are most commonly discussed in this context
are the abundance and the spatial clustering.  The model predictions
depend sensitively on the cosmology and on the value of $\sigma_8$,
the rms density fluctuations on the scale of $8$ $h^{-1}$ Mpc. (Here and
throughout this paper, $h$ is the present-day Hubble constant in
units of 100 km/s/Mpc.)  Comparisons between observations and model
predictions have been used  to place constraints on
cosmological parameters (Strauss {\it et al} 1995, Eke, Cole \& Frenk
1996, Viana \& Liddle 1996; Mo, Jing \& White 1996, Borgani {\it et
  al} 1997, De Theije, Van Kampen \& Slijkhuis 1998, Postman 1998 ).

It has  long been known that clusters of galaxies are much more 
strongly clustered than galaxies (see, for example, Hauser \& Peebles 1973
and review by Bahcall 1988).  The two-point correlation function for the
clusters is roughly a power law: $\xi_{cc}(r)=(r/R_0)^{-1.8}$.  
Bahcall \& West (1992) argue that the correlation length, $R_0$, obeys
the scaling relation 
\begin{equation}\label{bahcall-scaling}
R_0\approx 0.4D_c,\ \ 
20h^{-1}\,{\rm Mpc} < D_c < 100h^{-1}\,{\rm Mpc},
\end{equation}
where $D_c\equiv n_c^{-1/3}$ is the mean intercluster separation and
$n_c$, is the mean space density of clusters.  The combined set of
results based on the analysis of the spatial clustering of an X-ray
flux-limited sample of clusters (Lahav \etal~1989; Romer \etal~1994,
Abadi, Lambas \& Muriel 1998), of clusters containing cD galaxies
(West \& Van den Bergh 1991), of richness class R $\geq 0$, R $\geq
1$, R $\geq 2$ Abell clusters (Peacock \& West 1992; Postman, Huchra
\& Geller 1992), and of the  cluster samples extracted from
the APM Galaxy Survey (Dalton \etal~1992) and Edinburgh-Durham
Southern Galaxy Catalogue (Nichol \etal~1992) all give results that
are roughly consistent with the above scaling relation.

However, on scales greater than $D_c\approx 40 h^{-1}\,{\rm Mpc}$, the
evidence in favour of the scaling relation hinges just on the analyses of
the R $\geq 1$ and R $\geq 2$ Abell cluster samples, which give $R_0
\approx 21 h^{-1}\,{\rm Mpc}$ for $D_c \approx 55 h^{-1}\,{\rm Mpc}$
and $R_0 \approx 45 h^{-1}\,{\rm Mpc}$ for $D_c \approx 94
h^{-1}\,{\rm Mpc}$, respectively.  Several authors (e.g.~Sutherland
1988; Dekel \etal~1989; Sutherland \& Efstathiou 1991) have suggested
that these correlation lengths have been biased upward by the
inhomogeneities and projection effects in the Abell catalogue.
However, this suggestion has been rejected by, for example, Jing,
Plionis \& Valdarnini (1992) and Peacock \& West (1992). More
recently, Croft \etal~(1997) have analyzed the correlation properties
of a sample of ``rich'' APM clusters and find that the cluster
correlation length saturates at $R_0 \approx 15h^{-1}\, {\rm Mpc}$
($R_0 \approx 20 h^{-1}\, {\rm Mpc}$ if the analysis is done in
redshift-space --- see Croft \etal~1997) for $D_c > 40 h^{-1}\,{\rm
Mpc}$.  The controversy regarding the correlation length of rich
clusters: i.e. if the $R_0$ vs $D_c$ flattens at large scales is as of
yet still unresolved.

In an effort to resolve this issue, several authors (e.g.~Bahcall \&
Cen 1992; Watanabe \etal~1994; Croft \& Efstathiou 1994, 1997; Walter
\& Klypin 1996, Eke \etal~1996) have turned to large numerical
simulations.  Bahcall \& Cen (1992) investigated the cluster
correlation properties in large N-body simulations of the standard CDM
model (SCDM) and two low-$\Omega_0$ models ( $\Omega_0$ is the density
parameter), one spatially flat and one open.  They claim to find a
linear relation between $R_0$ and $D_c$ over the range $30h^{-1}\,{\rm
Mpc} < D_c < 95h^{-1}\,{\rm Mpc}$ in all the models but that only in
the low-density models is the $R_0$-$D_c$ relation steep enough to be
consistent with the suggested scaling relation
(\ref{bahcall-scaling}).  More recent works (Croft \& Efstathiou 1994,
Watanabe \etal~ 1994) have confirmed that the SCDM model is
incompatible with the observed degree of clustering on all scales and
for all normalizations.  However, no general agreement was reached on
the clustering strength at large scales for the other models
investigated.

In summary, apart from the general agreement that the SCDM model fails
to account for the observed cluster correlations, results obtained
from the numerical studies, due to lack of consistency, have been
singularly unhelpful in resolving the cluster correlation controversy.

If cluster correlations are going to be used to constrain models of
structure formation and place limits on the values of the fundamental
cosmological parameters, it is important to understand why these
numerical studies give such discrepant results.  This  is a
necessary step before a meaningful comparison between theoretical
(numerical) predictions and observations is possible.  There are
several factors that can affect numerical results and cause the
discrepancy described above.  Among these are differences in the mass
and force resolution of the simulations as well as the overall volume
of the simulations. Rich clusters tend to be rare objects and, therefore,
simulation studies of the properties of such objects must necessarily
span large cosmological volumes. Often, computational limitations
require that such simulation studies compromise on the resolution
(mass and/or force).  However, this can have serious effects on the
results. Watanabe \etal~(1994), have shown that degrading
the mass resolution tends to bias the correlation lengths downward.
Consequently, there is a definite need for analysis of a sample of
simulated clusters extracted from a simulation with high mass and
force resolution, large number of time steps, and covering a
sufficiently large cosmological volume.

In addition to differences in resolution and size, there is the
issue of how to identify clusters in the simulations.  Bahcall \& Cen
(1992), Watanabe \etal~(1994) and Croft \etal~(1994; 1997) used
different algorithms to identify clusters in their simulations.
Using a $\Omega=1$ SCDM model  Eke \etal~(1996b)  explored
the possibility that different algorithms could indeed give different
results.  They identified and ranked the clusters in their simulations
in various different ways, and found that each algorithm/selection
criteria imprints its own particular set of biases on the cluster
sample; for a fixed value of $D_c$, the clustering length can vary up
to a factor of $\sim$ 1.5.

In this paper, we report on our analysis of cluster correlations in
simulations of both critical density ($\Omega=1$) and open
($\Omega_0=0.3$ and $\Omega_0=0.4$) CDM cosmogonies and use our
results to explore the questions raised above.  As described in the
next section, both the force and  mass resolution of our
simulations are better than those of  previous studies. Moreover,
our simulated volumes are comparable with the Sloan Digital Sky Survey
(Loveday 1998) and are larger than the 2dF survey (Colless \& Boyle
1998).

We investigate the present-day abundances and temporal evolution of
the abundances in the three CDM models.  Specifically, we are
interested in testing the validity of the widely-used analytic
Press-Schechter expression for the cluster mass function.  The
combination of the present-day abundance of clusters and the rate at
which the abundance evolves as a function of time place strong
constraints on $\Omega_0$ and $\sigma_8$.  (White, Efstathiou \& Frenk
1993; Viana \& Liddle 1996; Eke, Cole \& Frenk 1996).  Since real
clusters of galaxies are the product of non-linear gravitational and
gas dynamical processes, the most direct way of constraining the range
of $\Omega_0$ and $\sigma_8$ is to carry out large-scale numerical
simulations of different models that have the necessary dynamical
range and include (poorly known) gas--stellar physics, then
``observe'' the resulting model universe and compare the simulated
observations with the real ones. Computationally, this route is
prohibitively expensive at present.  A more economical approach
involves using the analytic Press-Schechter (PS) formalism (Press \&
Schechter 1974; Bond \etal~1991) to compute the cluster mass function,
map the mass function into an abundance distribution as a function of
the observable parameter, and then compare the latter to observations
in order to determine the appropriate values of $\Omega_0$ and
$\sigma_8$.  Setting aside the uncertainties in the correspondence
between mass and an observable quantity, the validity of the analytic
approach rests entirely on the assumption that the PS formalism yields
an accurate description of the cluster mass function.  The analytic
expression for the cluster mass function has  been
extensively tested against numerical simulations in the past (see, for
example, Carlberg \& Couchman 1989; Lacey \& Cole 1994; Klypin
\etal~1994, Cole, Weinberg, Frenk \& Ratra 1997, Cen 1998) and most
studies have found a good agreement between the analytical and the
numerical results.  However, there have also been some interesting
claims to the contrary.  Gross \etal~(1998), for example, have drawn
attention to a discrepancy between the PS predictions and numerical
results at small masses, and Bertschinger \& Jain (1994) claim that
the PS mass function systematically underestimates the number density
of high mass halos.  Estimates of parameters such as $\Omega_0$ and
$\sigma_8$ are usually derived from fitting the analytic cluster mass
function to the observed distribution.  If the PS mass function is
indeed failing at the high mass end and this failure is not taken in
account, it can affect the determinations of $\Omega_0$ and $\sigma_8$

 In this paper our aim is to determine the halo mass function on group
and cluster scales in our set of simulations, and use these to assess
the reliability of the analytic PS mass function.  Each of our volumes
contains several hundred `Coma--like' clusters at the present
time. This, in conjunction with our high mass and force resolution,
allows us to map out the cluster mass function to  high precision out to
$z\sim 1$, i.e. over a larger redshift range than previously possible.

 The lay-out of the paper is as follows:  In \S 2, we discuss our
numerical simulations and the procedure for constructing 
cluster catalogs.  We use these catalogs to
extract  the cluster mass function and to study their spatial
correlation properties.  In \S 3, we present the results of our
correlation analyses and in \S 4, we discuss the cluster mass
function. In both sections we compare our numerical results to
analytical approximations.  Adopting a simple mapping between mass
and X--ray temperature, we transform our numerical mass function into
a temperature function and highlight the main differences between this
temperature function and the one based on the standard Press-Schechter 
mass function.  Finally, we summarize our results and briefly discuss their 
relevance for future cosmological tests in \S 5.

\section{Numerical simulations and cluster selection}

We have simulated structure formation within a periodic cube of
comoving length $L=500h^{-1}$ Mpc for two ``fiducial'' cosmological
models: A critical density cold dark matter models ($\Omega_0=1$,
$h=0.5$ with $\sigma_8=1.0$ at $z=0$ ( hereafter we refer to the
$\sigma_8=0.7$ output as SCDM07 and to the $\sigma_8=1$ as SCDM10,
respectively. Of course each output of the SCDM run can be rescaled to
a different z changing the present day $\sigma_8$ normalization) and
an open ($\Omega_0 = 0.3$, $h=0.75$, $\sigma_8=1.0$ at $z=0$ ---
hereafter referred to as O3CDM) cold dark matter model.  The $z=0.58$
output of O3CDM simulation can, with appropriate rescaling, be
identified as an $\Omega_0 = 0.4$, $h=0.65$, $\sigma_8=0.79$ CDM
simulation (hereafter referred to as O4CDM) of comoving length
$L=433.3\;h^{-1}$ Mpc. The same set of simulations was used by Szapudi
\etal~(1998) to study the higher order correlation properties of
galaxies.  The initial conditions were set using the Bardeen {\it et
al} (1986) transfer function for CDM. The simulations were computed
using {\em PKDGRAV}, a parallel treecode that allows for periodic
boundary conditions and individual time steps (Stadel \& Quinn, in
preparation). These are among best studied cosmological models (Davis
\etal~1985, Jenkins \etal~1997), our choices for the normalization
($\sigma_8$) of the open models correspond roughly to those inferred
from the present-day cluster abundance (see, e.g., Eke et al. 1996 and
references therein), while we analyzed the SCDM simulation data on a
$\sigma_8$ range that goes from $\sigma_8$ = 1 (roughly COBE
normalized) to $\sigma_8$ = 0.35 (corresponding to z $=$ 1.85 for a
$\sigma_8$ = 1 at z = 0 model, and to z $=$ 0.43 for a cluster
normalized SCDM universe with $\sigma_8$ = 0.5 at z = 0) . A cubic
spline force softening of 50$h^{-1}$ kpc (43$h^{-1}$ kpc for O4CDM)
was used so that the overall structure of clusters could be resolved.
Accurate forces were maintained by using a cell opening angle of
$\theta < 0.8$ (or better at high z) and expanding the potentials of
cells to hexadecapole order.  Timesteps were constrained to $\Delta t
< 0.3 \sqrt{\epsilon/a}$, where $\epsilon$ is the softening length and
$a$ is the magnitude of the acceleration of a given particle.  See
Quinn \etal~(1997) for a discussion and tests of this timestep
criterion.  In each run 47 million particles were used, arranged on a
360$^3$ grid.  Each run took several hundred hours on 256 nodes of a
Cray T3E supercomputer, and about a thousand timesteps.  The particle
mass is $7.4\times 10^{11}\eta\Omega_0 h^{-1}\,{\rm M}_\odot$, where
$\eta=1$ for SCDM and O3CDM models and $\eta=0.65$ for
O4CDM. Simulations were started at z $=$ 49.  The extremely large
volumes simulated, coupled with a reasonable mass resolution and the
very good force resolution made possible by the use of a treecode,
allow us to study in detail the evolution of structures ranging in
size from groups of galaxies, made up of several tens of particles
each, to very rich clusters that contain a few thousand particles.  In
our analyses, we only consider halos consisting of 64 particles or
more.  This is a stricter constraint than used in most previous work,
and was imposed to ensure that our results were not influenced by
small-number effects. Finally, we verified that in these simulations
both the initial and present--day power spectrum were in close
agreement with theoretical expectations ( see Peacock \& Dodds 1996).

\subsection{Cluster Identification and Selection}

Theoretical treatments generally define virialized halos at a given
epoch as structures with a mean density averaged over a sphere of
$\sim 200$ times the critical density at that epoch (see, for example,
 Lacey \& Cole 1994 and references therein).  The mass contained
in the sphere is taken to be the mass of the halo and the radius of the
sphere is usually identified as the virial radius of the halo.

In numerical simulations, halos are identified using a variety of
schemes.  Of these, we have chosen to use two that are available in
the public domain:
FOF\footnote{http://www-hpcc.astro.washington.edu/tools/FOF/} (Davis
\etal 1985), and
HOP\footnote{http://www.sns.ias.edu/eisenste/hop/hop.html} (Eisenstein
\& Hut 1998).  These schemes are discussed in the next subsection.
Other halo finders that are often used in literature to find
virialized halos are DENMAX (Gelb \& Bertschinger 1994), the
``spherical overdensity algorithm '' or SO, that finds spherically
averaged halos above a given overdensity (Lacey \& Cole 1994) and the
scheme recently developed by Gross {\it et al} (1998).  The algorithms
that we opted to use are those in the public domain and hence, in
common use.  We felt that it was important to ascertain the extent to
which these schemes may bias our results.

\subsection{ Friend of Friends: FOF}

The FOF algorithm (Davis \etal~1985) is one of the most widely
used. It is based on a nearest neighbor search.  The main advantages
of this algorithm are its simplicity and the lack of assumptions about
the shape of halos.  In this scheme, all particle pairs separated by
less than $b$ times the mean interparticle separation are linked
together.  Sets of mutually linked particles form groups that are then
identified as dark matter halos.  In the present study, we adopted the
linking length that Lacey \& Cole (1994) arrived at  to
identify virialized halos with mean densities of $\simeq 200$ times
the critical density at the epoch under consideration.  The linking
length is $0.2\Omega(z)^{-1/3}$ times the mean comoving interparticle
separation.  Moreover in the low $\Omega$ models, the scaling of the
linking length as a function of redshift was further modified as the
mean halo density associated with virialization is a function of
redshift (see for example, Kitayama \& Suto 1996).
The resulting halos also have a mass--radius relation that agrees
reasonably well with the theoretical relation for virialized halos
(see Lacey \& Cole 1994; also, Eke, Cole, Frenk \& Navarro 1996).  The
objects identified by the FOF algorithm are the kinds of objects that
the PS formalism refers to (apart from the lack of spherical symmetry)
and therefore, we should be able to make a meaningful comparison
between the distribution of halos in the simulations and the PS
distribution.
Several authors have reported the tendency of FOF to link together 
close binary systems of similar mass especially if the two happen to
be loosely connected by a bridge of particles. This pathology can, in 
specific cases (see for example Governato {\it et al.}~1997), give rise
to biased results.  We have verified that our results are largely unaffected
by this problem.

We note that in their study, Lacey \& Cole (1994) compared the 
properties of the halo population defined by FOF and with those of a 
sample generated using the SO algorithm. They 
found that at least over the mass range that they could probe using their 
simulations, the two algorithms gave very similar results.

\subsection{HOP}

HOP is a recently introduced algorithm (Eisenstein \& Hut 1998) based
on an hybrid approach.  The local density field is first obtained
by smoothing the density field with an SPH--like kernel  using the
{\it n}  nearest neighbours (we used 16). The particles above a
given threshold are linked with their highest density neighbors until,
after several ``hops'', they are connected to the one particle with
the highest density within the region above the threshold.  All
particles linked to the local density maximum are identified as a
group.  Like FOF, HOP is well suited to identifing virialized structures
once the density threshold is specified to be the local density at the
virial radius.  Eisenstein \& Hut (1998) claim good
agreement with the FOF method at masses above the smoothing
scale. However, HOP can be tuned to separate binary halos---binary
systems loosely connected by one--dimensional particle
bridges---thereby avoiding the (rare) FOF pathology.

\subsection{FOF vs. HOP}

\begin{figure}
{\epsfxsize=8.truecm \epsfysize=10.truecm 
\epsfbox[40 130 560 700]{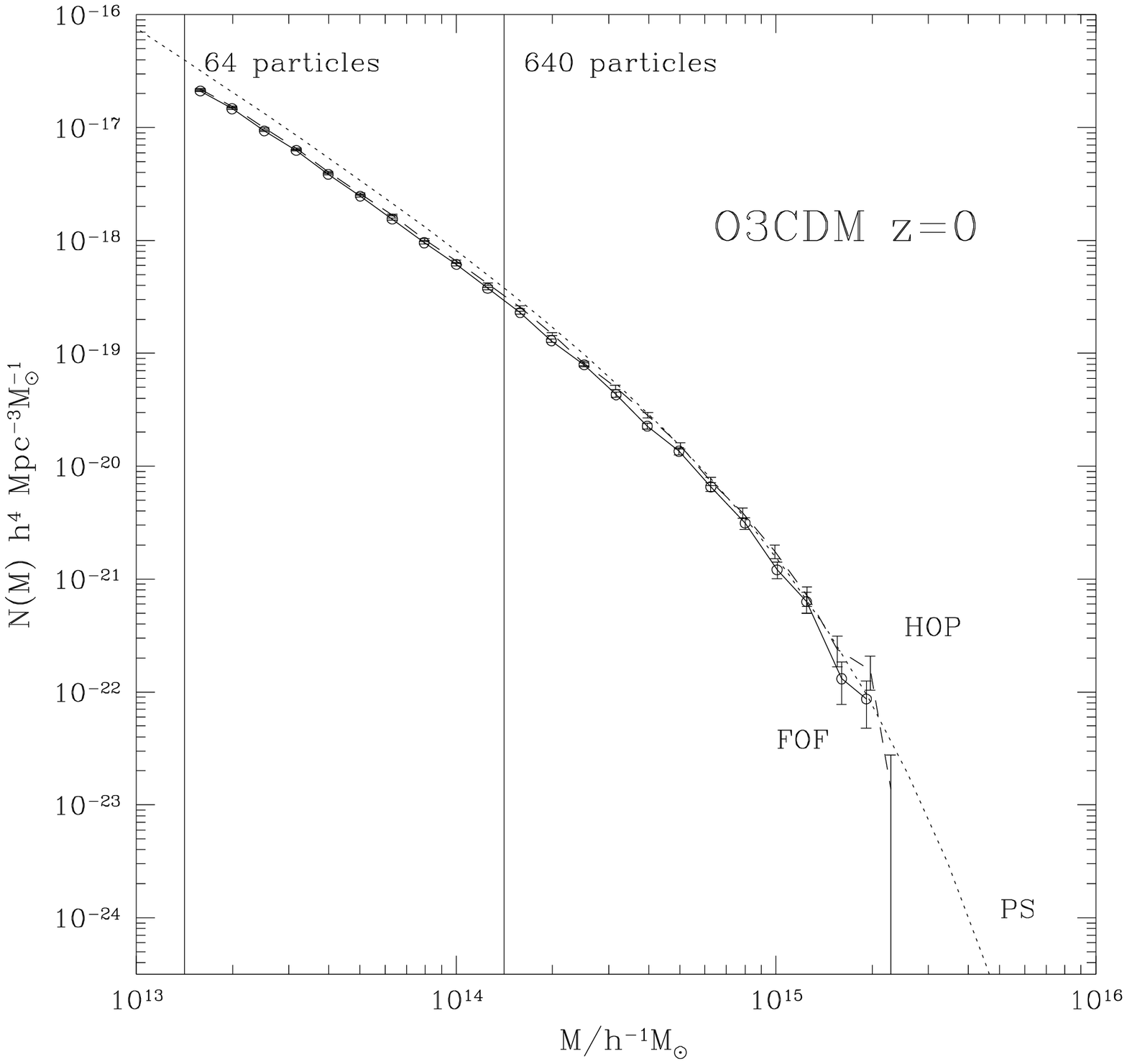}}
\caption{ {Differential mass function (number density per unit mass)
of groups and clusters extracted from the $z=0$ O3CDM simulation
volume using FOF (continous line) and HOP (dot--dashed line)
algorithms.  The dotted line shows the analytic Press-Schechter
prediction for the mass function.
\label{hfcomp}}
}
\end{figure}
In Figure \ref{hfcomp} we show the mass functions obtained by applying
the two halo finders described to the O3CDM run.  The results for SCDM
runs are qualitatively similar.  The FOF and HOP mass functions agree
quite well over the entire mass range probed, with most massive HOP
clusters showing a systematic offset of about 7\% toward larger
masses. This offset can be easily adjusted changing {\it b} or the
density threshold for HOP. However, as discussed in the above
paragraph the parameters used are the most physically
meaningful and the small offset is a measure of the kind of biases
that you get using different halo finders.

We have found this general agreement to hold for different models and at
all redshifts. This result, coupled with results of Eke {\it et al.}'s 
(1996) comparison of the FOF and SO algorithms, strongly indicates that
regardless of the actual details of the scheme used to identify the
halos, if the resulting halos are independent virialized entities then
the statistical properties of the halo populations will be very similar.
Figure \ref{hfcomp} also shows the PS prediction as a comparison.  We defer
the comparisons of the theoretical curve to the numerical results in \S 4.

\begin{figure*}
\centerline{\hbox{
\psfig{figure=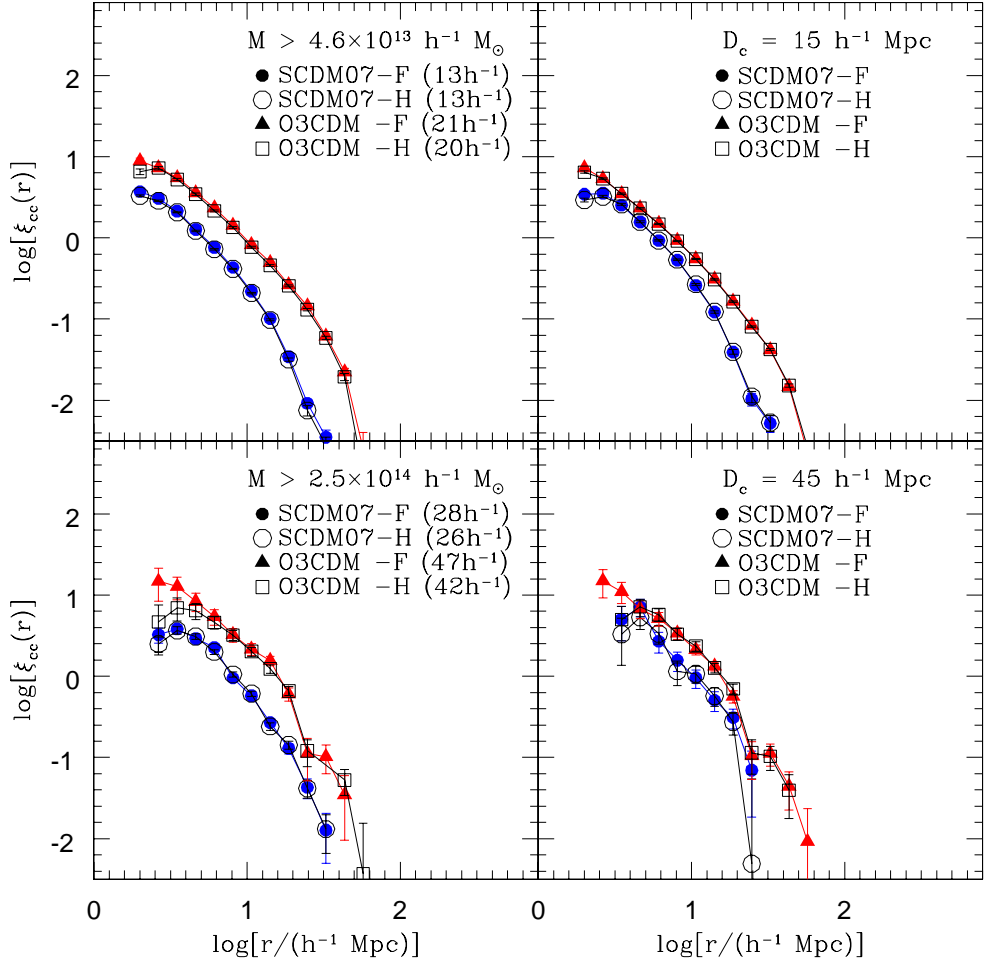,height=15cm,width=15cm,angle=0}
}}
\caption{ Real-space correlation functions of clusters extracted from
the SCDM07 output and O3CDM at z$=$0 using either (F)OF or (H)OP
algorithms.  The two left panels show the correlation functions for
clusters with masses greater than the specified threshold.  The
numbers in the parenthesis are the $D_c\equiv n_c^{-1/3}$ values for
the cluster samples.  The two right panels show correlation functions
for samples with the cluster number density given the specified value
of $D_c$.
\label{fof-uskid-hop}
}
\end{figure*}

\section{THE TWO-POINT CLUSTER CORRELATION FUNCTION}

The output of our cosmological simulations was processed using the two
halo identification algorithms (HOP \& FOF) described in the preceding
section.  We ordered the lists according to halo mass and then
generated cluster catalogs by applying a fixed lower mass cutoff.  We
also generated cluster catalogs based on the ordered list with a
specific number density of clusters (labelled by the corresponding
value of $D_c$).

For each cluster catalog, we compute the real space two-point correlation
function using the direct estimator:
\begin{equation}
\xi_{cc}(r)={N_p(r)\over n_c^2 V (\delta V)} -1,
\end{equation}
where $N_p(r)$ is the number of cluster pairs in the radial bin of volume
$\delta V$ centered at $r$, $n_c$ is the mean space density of the cluster
catalog and $V$ is the volume of the simulation.  We use all the clusters in 
our catalogs, taking advantage of the periodic boundary conditions.

The 1$\sigma$ error bars for the correlations are estimated using the formula
\begin{equation}
\delta\xi_{cc}(r)=\frac{3}{2}\frac{1}{\sqrt{N_{cc}(r)}}
\left(1+\xi_{cc}(r)\right),
\end{equation}
where $N_{cc}$ is the number of distinct cluster pairs in the radial bin at
$r$.  
We have increased the size of the Poisson error bars by $50\%$ 
because these errors do not take into account clustering and 
so are likely to underestimate the true errors 
(Croft \& Efstathiou 1994; Croft \etal 1997).

The correlation functions are not well described by a single power-law 
over the entire range of pair separations sampled.  
To estimate the correlation length, we fit a functional form 
\begin{equation}
\xi_{cc}(r)=\left({r\over R_0}\right)^{-\gamma}
\end{equation}
over the range $ 4.5h^{-1}\,{\rm Mpc} < r < 25h^{-1}\,{\rm Mpc}$, 
which brackets the point where $\xi_{cc}=1$.  
We estimate the value of $R_0$
by both fixing the value of $\gamma$ in the above equation to $-1.8$
(see equation \ref{bahcall-scaling}) and by allowing $\gamma$ to be a 
free parameter.  Since, the fit is done over a restricted range in $r$,
both schemes yield similar values of $R_0$.

\begin{figure*}
\centerline{\hbox{
\psfig{figure=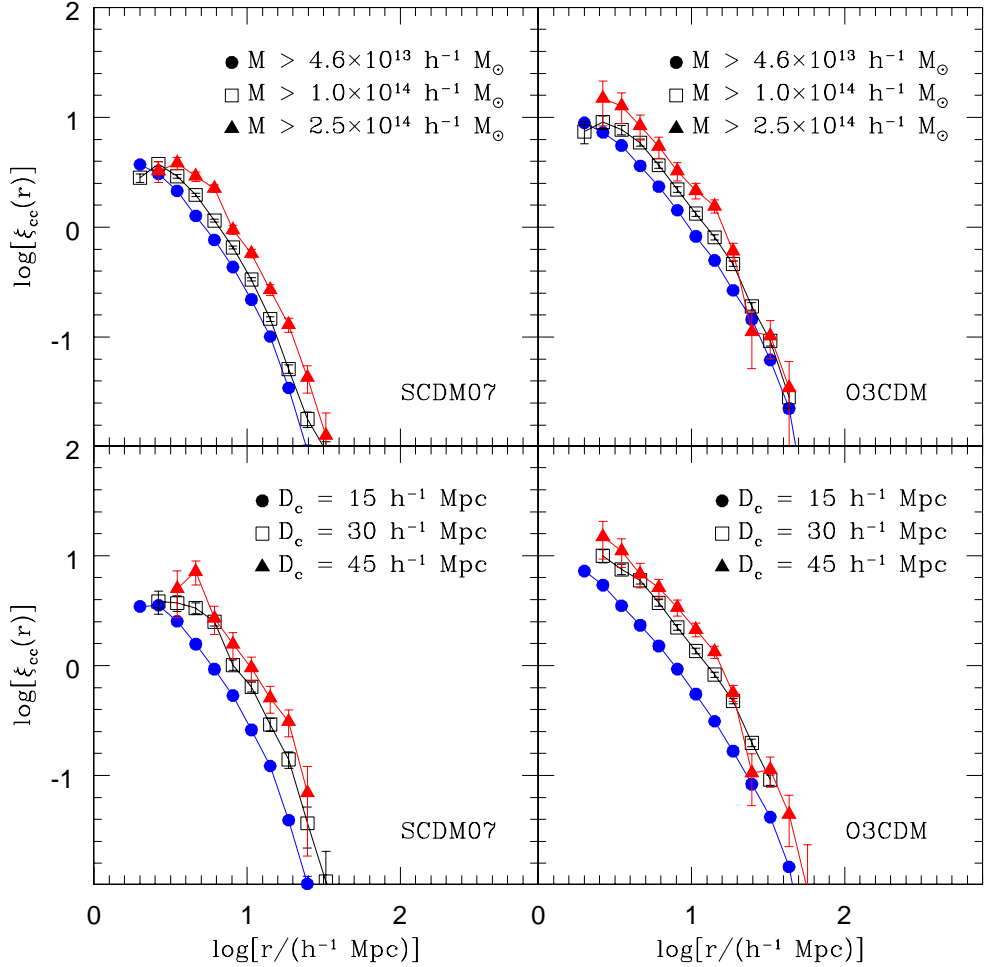,height=15cm,width=15cm,angle=0}
}}
\caption{
Real-space correlation functions of clusters samples defined by either 
imposing different mass thresholds (top two panels) or by demanding 
that the sample clusters have some predefined number density (bottom
two panels).  The clusters have been extracted from the SCDM07 output and
O3CDM at z$=$0 using the FOF algorithm. 
\label{mass-rich}
}
\end{figure*}

In Figure \ref{fof-uskid-hop}, we show real-space correlation
functions of clusters in catalogs defined by two different lower mass
thresholds ($M_{cut}=4.6\times10^{13}h^{-1}\;M_\odot$ and
$2.5\times10^{14}h^{-1}\; M_\odot$) and two different cluster
abundance requirements ($D_c=15h^{-1}$ Mpc and $40h^{-1}$ Mpc).  The
clusters are extracted from the simulations using either the FOF or
HOP algorithms. Figure 2 shows the results for clusters extracted from
the SCDM07 output and the O3CDM simulation at z$=$0; the clustering
trends of O4CDM and SCDM10 clusters are  the same.

At both low and high mass thresholds, the correlation functions of FOF
and HOP clusters are virtually identical, especially in the range $
4.5h^{-1}\,{\rm Mpc} < r < 25h^{-1}\,{\rm Mpc}$. The abundances of FOF
and HOP clusters (or equivalently, their $D_c$ value) are also the
same, as expected from results shown in Figure 1.  As the mass
threshold increases, or the number density is decreased, the clustering
amplitude increases, but the shape of the correlation function remains
the same. This reflects the fact that massive, rare peaks tend to be
more strongly clustered in all CDM models.  This result is shown in
Figure \ref{mass-rich}.

Given the good match between the halo catalogs we will mainly discuss
results for the FOF clusters. Unless specified, results for FOF
clusters hold for the HOP clusters as well.

\subsection{Cosmology and Normalization of the Mass Power Spectrum}

\begin{figure}
{\epsfxsize=7.5truecm \epsfysize=15.truecm 
\epsfbox[45 170 320 690]{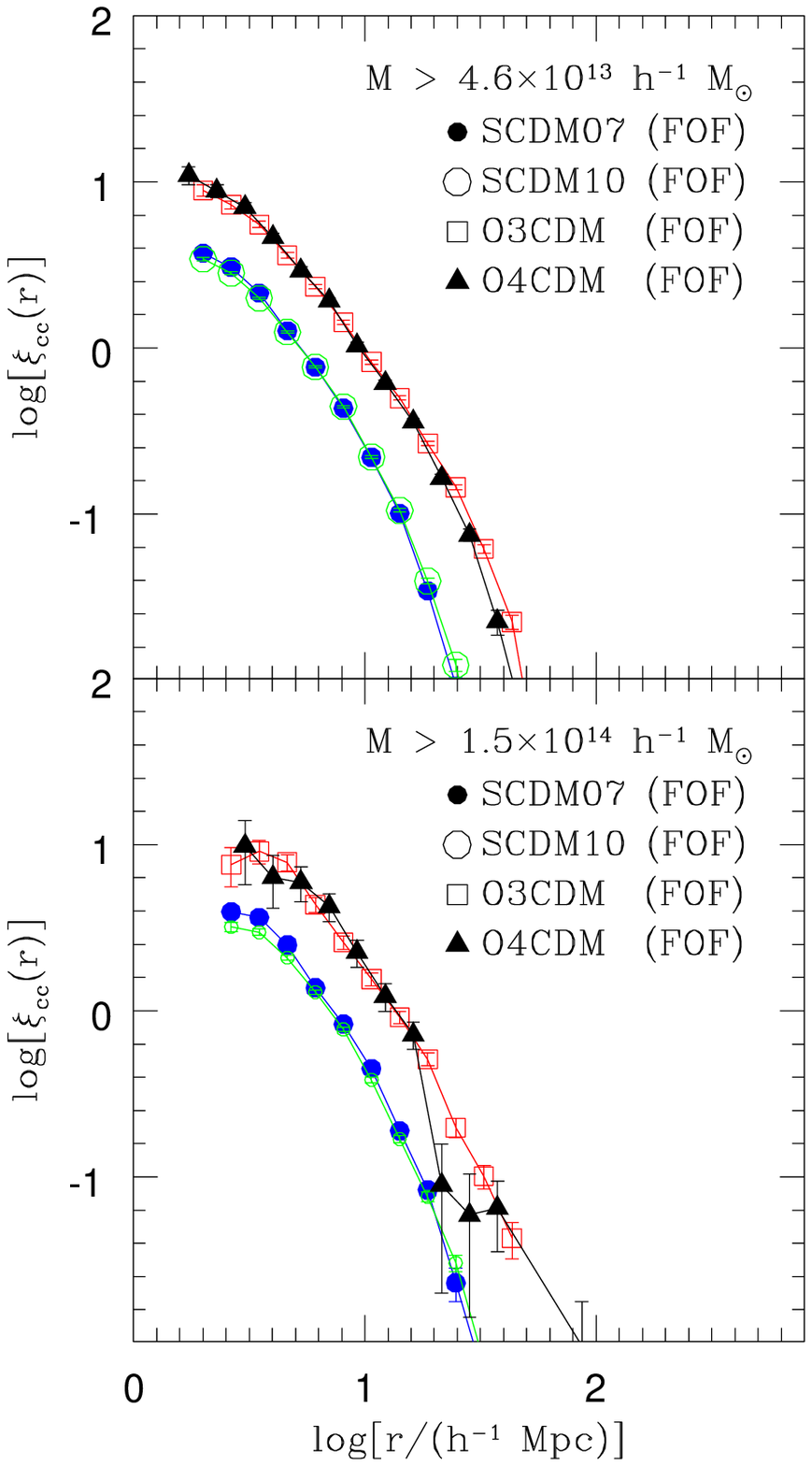}}
\caption{
Real-space $z=0$ cluster correlation functions extracted from our 
simulations (SCDM07, SCDM10, O3CDM, O4CDM) using the FOF algorithm.
\label{norm-cosmo}
}
\end{figure}

The real-space $z=0$ correlation functions of FOF cluster samples from the
various  simulations are compared in Figure \ref{norm-cosmo}.  
Considering the SCDM10 and SCDM07 as two z$=$0 outputs  results in 
the model with the higher amplitude (SCDM10) developing structure on group 
and cluster scales at an earlier epoch, having a higher density of very 
massive halos and a more strongly clustered mass density field at the present 
epoch. 

In spite of the above mentioned differences, the cluster correlations for 
the SCDM models with different normalisations are virtually identical in 
shape and amplitude for cluster samples with both high as well as low mass
thresholds.  In the case of massive clusters, this has been previously 
noted by both Croft \& Efstathiou (1994) and Eke \etal~(1996).  Since
for the SCDM models, studying the changes (or lack thereof) in the 
correlation functions due to variations in the normalization of the 
amplitude of the primordial density fluctuations is equivalent to studying 
the evolution of the clustering property as a function of time, we defer
the discussion of the above-mentioned until \S3.3.

The correlations for the two open models are also very similar to each
other, both in shape and amplitude.  These two models differ not only
in their values of $\Omega_0$ and $h$ but also in the normalization of
the amplitudes of the primordial mass fluctuations as defined by
$\sigma_8$.  The two OCDM models do, however, have the same value of
$\Omega_0 h^2$. Since it is
this parameter that defines the position of the peak in the CDM power
spectrum characterizing the initial Gaussian random fluctuations in
density field, it is perhaps not surprising that the cluster
correlations are similar.

In comparison to the cluster correlations in a critical universe, the
OCDM cluster correlation functions have a significantly higher
amplitude.  This occurs because the peak in the power
spectrum for the OCDM models is displaced towards larger scales and
therefore, for a similar value of $\sigma_8$, the OCDM models have more
power on large scales than the SCDM models.

\subsection{Redshift Evolution}

\begin{figure}
{\epsfxsize=7.5truecm \epsfysize=12.truecm 
\epsfbox[45 170 320 690]{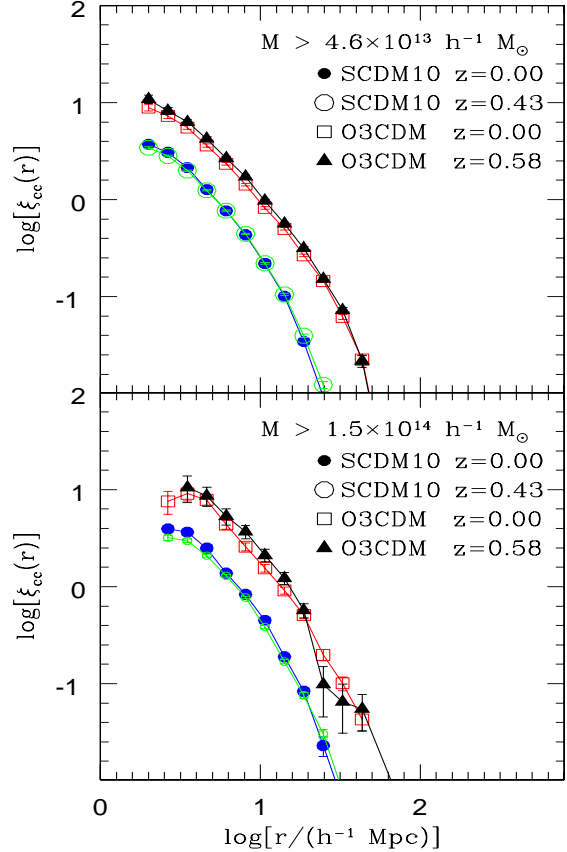}}
\caption{
Correlation functions of SCDM10 and O3CDM clusters computed at two different 
epochs and plotted in comoving coordinates.  
\label{evolv}
}
\end{figure}

In Figure \ref{evolv}, we plot the present-day and $z=0.43$, $z=0.58$
cluster correlations, in comoving coordinates, for two cluster samples
defined as ($M>4.6\times 10^{13}h^{-1}\; M_\odot$ and $M>1.5\times
10^{14}h^{-1}\; M_\odot$) drawn from the SCDM10 and O3CDM models
respectively.

Before we discuss the results, let us consider what is expected.
Given a sample of halos with masses greater than some threshold
$M_{cut}$, the correlation function of the halos can be related to
that of the total mass distribution via the bias
parameter:\hfill\break $\xi_{CC}(r; M >M_{cut} ) =
b^{2}_{eff}(M_{cut}) \xi_{\rho\rho}(r)$.  At a given epoch the bias
parameter becomes larger as $M_{cut}$ is raised, as we have already
shown.  For a fixed $M_{cut}$ and a critical universe, the bias
parameter is expected to decrease asymptotically to unity as a
function of time (Tegmark \& Peebles 1998) as the underlying mass
distribution becomes more clustered.  The time evolution
of $\xi_{CC}(r; M >M_{cut} )$ depends on the competition between these
two trends.

Turning to Figure \ref{evolv}, we note that for both low and high mass
thresholds, there are no significant differences between the comoving
correlation functions at $z=0$ ($\sigma_8$=1) and $z=0.43$ for the SCDM
clusters.  This implies that over the mass and redshift ranges
considered here, the rate of increase in the clustering of the mass
distribution is closely matched by the rate at which the bias
parameter decreases.

For the O3CDM model, the correlation function at the earlier epoch has 
a slightly higher amplitude for both low and high threshold samples. The
comoving correlation length at $z=0.58$ is a factor of 1.1--1.2 greater.

{\it To summarize, the comoving group/cluster correlation functions
are either constant or change very little over the redshift range $0 <
z < 0.5$ and in proper coordinates, the group/cluster correlation
length decreases with increasing redshift over the redshift range
studied.  In the SCDM case, this decrease is given by $R_0\propto
(1+z)^{-1}$ and in the O3CDM model, by $R_0\propto (1+z)^{-0.86}$.}

\subsection{Comparison with analytic calculations}

\begin{table}
\begin{center}
\caption[dummy]{ The bias parameter for SCDM07 cluster samples.
$b_{1}$ is computed using the standard PS mass function. The second
column gives the mass cut in units of the characteristic mass, $M_{*}
= 4 \times 10^{13}h^{-1}\; M_{\odot}$.  The final column gives the
Lagrangian radius of the halo, which is the smallest separation where
 the assumtpions in the calculations are valid.  }
\begin{tabular}{ccccc}
\hline
\multicolumn{1}{l} {$M_{cut} h^{-1}\;M_{\odot}$} & 
\multicolumn{1}{l} {$M_{cut}/M_{*}$ } & 
\multicolumn{1}{l} {$b_{1}$ } &  
\multicolumn{1}{l} {$r_{L}h^{-1}$ Mpc } \\ 
\hline
        5.7e+14  &    16.2 &     3.3   & 7.9\\
        2.8e+14  &     7.0 &     2.4   & 6.2\\
        2.1e+14  &     5.2 &     2.2   & 5.7\\
        7.0e+13  &     1.7 &     1.5   & 3.9\\
\hline
\label{table:bias1}
\end{tabular}
\end{center}
\end{table}

\begin{table}
\begin{center}
\caption[dummy]{
The bias parameter for O3CDM cluster samples.
as in Table \ref{table:bias1}. 
In this case $M_{*} = 1.4 \times 10^{13}h^{-1}\; M_{\odot}$
}
\begin{tabular}{ccccc}
\hline
\multicolumn{1}{l} {$M_{cut} h^{-1}\; M_{\odot}$} & 
\multicolumn{1}{l} {$M_{cut}/M_{*}$ } & 
\multicolumn{1}{l} {$b_{1}$ } & 
\multicolumn{1}{l} {$r_{L}h^{-1}$ Mpc } \\ 
\hline
        2.7e+14  &    19.3 &    2.7  &9.2 \\
        1.9e+14  &    13.6 &    2.4  &8.2 \\
        1.1e+14  &     7.8 &    2.0  &6.8 \\
        9.0e+13  &     6.4 &    1.9  &6.4 \\
        1.8e+13  &     1.3 &    1.2  &3.7 \\
\hline
\label{table:bias2}
\end{tabular}
\end{center}
\end{table}

To date, most studies of cluster correlations have utilized numerical
simulations. Such numerical simulations are very expensive to
generate, a constraint that renders a systematic exploration of different
cosmological models impractical; it also makes it rather
difficult to explore and identify the general physical mechanisms
underlying the clustering properties of clusters and group halos viz a
viz that of the mass distribution. Consequently, various authors
(e.g.~Kaiser 1984; Bardeen \etal~1986; Mann \etal~1993;  Mo \&
White 1996; Mo, Jing \& White 1996, Catelan \etal 1998)
 have  developed analytic schemes to compute the cluster correlation function.

The first method to compute cluster correlation functions analytically
that we discuss  is based on the Press-Schecter formalism
and its extensions. This was originally developed by Cole \& Kaiser
(1989) and Mo \& White (1996) to derive a model for the spatial
correlation of dark matter halos in hierarchical models.  The
calculation consists of three steps (see Baugh \etal 1998).

$\bullet$ Compute the nonlinear power spectrum for the 
cosmology and $\sigma_{8}$ in question using the transformation of  
the linear power spectrum suggested by Peacock and Dodds (1996).

$\bullet$ Calculate an effective bias parameter, $b_{eff}$ for the dark matter 
halos above the specified mass cut as outlined by Mo \& White (1996).

$\bullet$ Fourier transform the nonlinear power spectrum to get 
the nonlinear correlation function of the mass, then multiply by the 
square of the halo bias factor, to get the real-space, nonlinear, 
cluster correlation function: 
$\xi_{CC}(r) = b^{2}_{eff} \xi_{\rho\rho}(r)$.

The cluster correlation function thus computed has been tested, against
 N-body results by Mo \& White (1996) and Mo, Jing \& White 
(1996) and is found to hold even in the mildly nonlinear regime where 
$\xi(r) > 1$ as long as $r > r_L$ where $r_{L} = \left(3M/4\pi 
\rho_{0}\right)^{1/3}$  is the Lagrangian radius of the dark matter halos
($r_L\sim 10h^{-1}$ Mpc for rich clusters of galaxies)
and $\rho_{0}$ is the present mean density.  Recently Jing (1998) has shown
that the Mo \& White formula systematically underpredicts  the bias of
low mass halos, but it is in good agreement with numerical simulations
in the mass range considered here.

The bias parameter for a dark matter halo that contains a single 
galaxy is given by the formula derived by Mo \& White (1996) and was
written down for any redshift in Baugh \etal~(1998):
\begin{equation}
b(M, z) = 1 + \frac{1}{\delta_c} \left[
\left(\frac{\delta_c}{\sigma(M)D(z)}\right)^2 -1 \right]
\label{eq:bias}
\end{equation}

Here $D(z)$ is the linear growth factor, normalized to $D(z=0)=1$, 
$\sigma(M)$ is the rms linear density fluctuation at $z=0$ and 
$\delta_{c}$ is the extrapolated  linear overdensity for 
collapse at redshift $z$.
This gives the bias factor for the halo when the clustering is measured 
at the same epoch that the halo is identified.

For a sample of halos with different masses, the effective bias is given
by
\begin{equation}
b_{eff}(z)={\int N(M,z)\; b(M,z)\; dM \over \int N(M,z)\; dM}
\end{equation}
where $N(M,z)\;dM$ is the number density of halos with mass $M$ in the sample.
For the cluster samples that we have constructed, $N(M,z)$ can either be 
set equal to the Press-Schechter mass function (with $\delta_c$ set to the
canonical value defining collapse for the cosmology under consideration)
or to the cluster mass function computed directly from the cluster catalogs.
However the results shown here are almost insensitive to this 
choice.

\begin{figure*}
\centerline{\hbox{
\psfig{figure=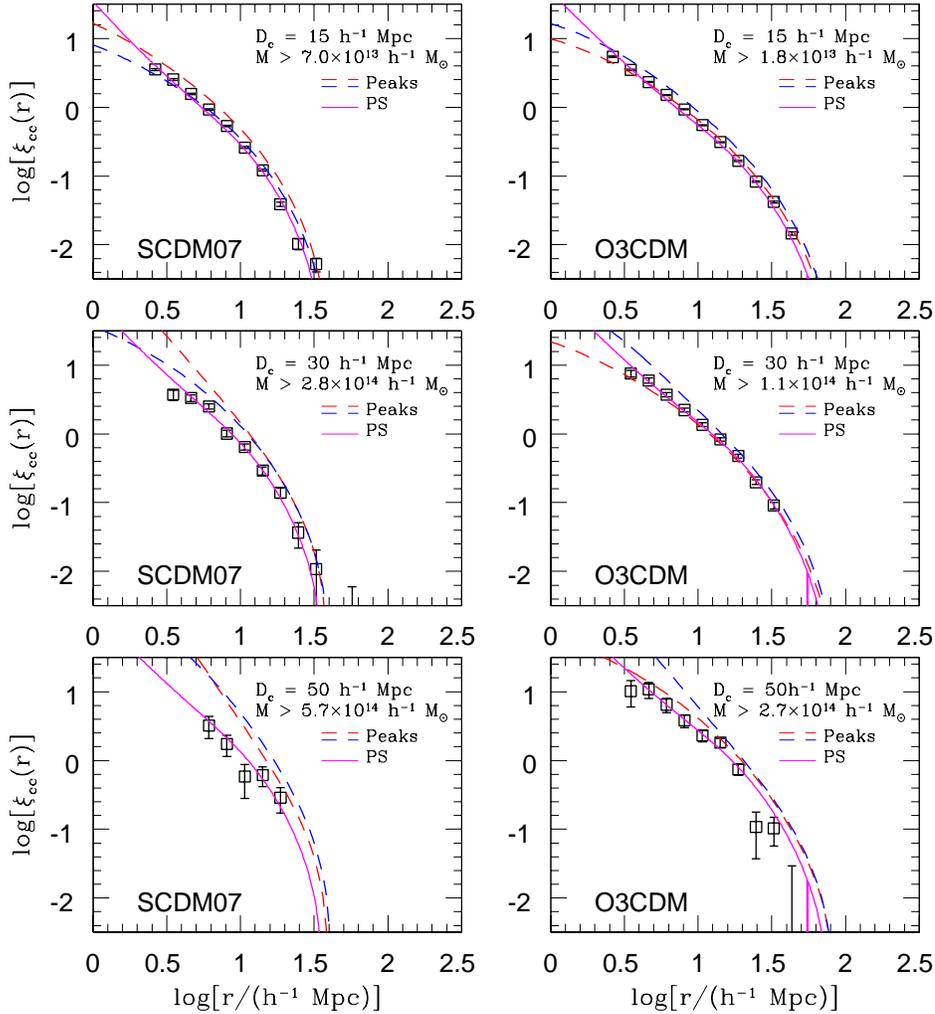,height=15cm,width=15cm,angle=0}
}}
\caption{
Analytic correlation functions compared against our numerical results.
The two dashed curves are the peaks-based correlation functions computed 
according to the prescription of Mann \etal~(1993).  The curve with the 
higher correlation amplitudes on small scales corresponds to 
$\delta_c=1.7$.  The other curve corresponds to $\delta_c=1.$.
The two solid curves are the PS-based correlation functions computed
as described in \S3.5.  The correlation
functions are computed assuming either the standard PS mass function or
the numerical cluster mass function (see \S4).  The two are very similar.
\label{analytic}
}
\end{figure*}

The effective bias parameters for samples whose correlation functions
are plotted in Figure \ref{analytic} are given in Tables
\ref{table:bias1} (SCDM07) and \ref{table:bias2} (O3CDM).  The mass
cuts applied correspond to halos of different rarity in the two
cosmologies; this is quantified by comparing the mass cut to the
characteristic mass $M_{*}$, which is defined later in \S 4.  All the
mass cuts considered correspond to objects that are greater than
$M_*$, and so these halos are biased tracers of the dark matter
distribution (Mo \& White 1996).  The cluster sample with the highest
mass cut for both O3CDM and SCDM07 is predicted to have a correlation
function that is $\sim 10$ times higher than that of the dark matter.

\begin{figure*}
\centerline{\hbox{
\psfig{figure=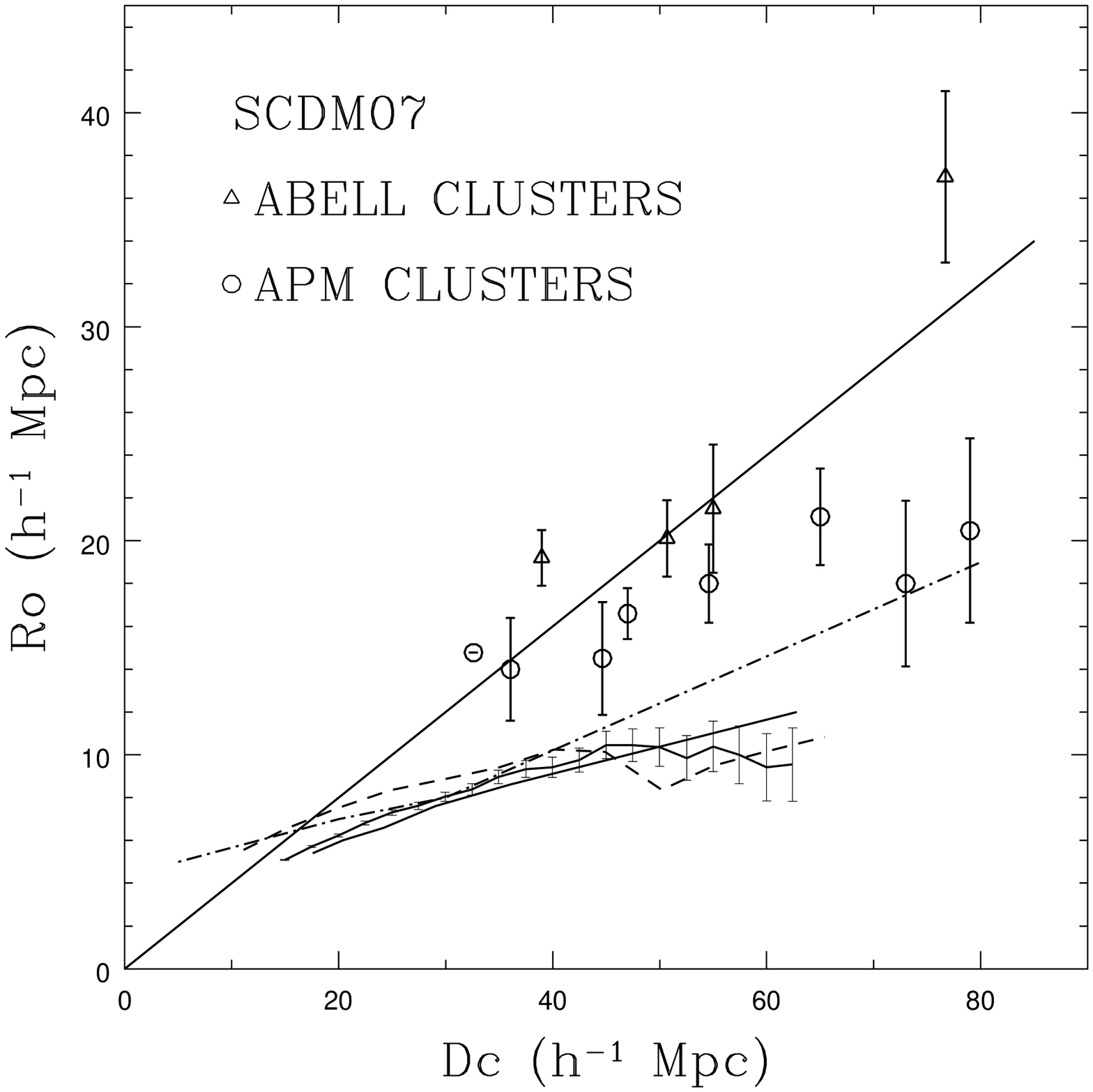,height=12cm,width=12cm,angle=0}
}}
\caption{ Cluster correlation length as a function of $D_c$, the mean
cluster separation, for clusters extracted from the SCDM07 output
using FOF (heavy solid curve).  The error bars show the 1$\sigma$
errors in $R_0$ derived from fitting the correlation functions with a
$-1.8$ power-law as described in the text.  The solid line
corresponds to the scaling relation between $R_0$ and $D_c$ (equation
\ref{bahcall-scaling}) advocated by Bahcall \& West (1992).  The
dot--dashed line shows the $R_0$--$D_c$ that Bahcall \& Cen (1992)
derived from their numerical study.  The short-dashed curve
corresponds to the numerical results of Croft \& Efstathiou (1994).
In addition, the open triangles show the results for R$\geq 0$, R$\geq
1$ and R$\geq 2$ Abell clusters (Bahcall \& Soneira 1983; Peacock \&
West 1992) and the open circles show the results for APM clusters
(Dalton \etal~1992; Croft \etal~1997).  The light solid curve is the
$R_0$--$D_c$ relation derived from analytic PS-based correlation
functions computed according to the prescription in Baugh \etal~(1998)
--- see \S3.5.
\label{neta-scdm}
}
\end{figure*}

The PS-based analytic correlation functions are shown in Figure
\ref{analytic} as solid curves.  There is little difference between
the correlation functions computed using the standard PS mass function
and the numerical mass function discussed in \S4.  The analytic
correlations are in excellent agreement with our numerical correlation
functions.  The agreement between the numerical and analytic results
is further confirmed by the match between the analytic and numerical
$R_0$--$D_c$ curves.  The analytic $R_0$--$D_c$ curve is plotted in
Figures \ref{neta-scdm} and \ref{neta-ocdm} as the light solid curve.

We consider next  the scheme developed by
Mann \etal~(1993).  This  is based on the method devised by 
Couchman \& Bond  (1988; 1989) that combines the theory of the 
statistics of peaks in Gaussian random fields with the 
dynamical evolution of the cosmological density field.

In this scheme, the time evolution of the density field is followed
using the Zel'dovich approximation (Zeldovich 1970).  At the epoch of
interest, a particular class of objects is defined by the pair $R_s$
and $\delta_c$.  These are, respectively, the smoothing scale that is
applied to the cosmological density field and the linearly
extrapolated amplitude of the density fluctuations at the time of
collapse.  Mann \etal~(1993) set the values of these two parameters by
choosing an appropriate value for $\delta_c$ ($\delta_c=1.686$,
 corresponds to collapse of spherical density perturbations in
an $\Omega_o=1$ universe) and then adjusting $R_s$ until the number
density of peaks with overdensities greater than $\delta_c$
corresponds to number density of objects under consideration.  Full
details can be found in Mann \etal~(1993).

In Figure \ref{analytic}, we  plot the correlation functions of
some of our cluster samples and show the corresponding analytic peaks-based 
correlation functions computed assuming $\delta_c=1$ and $\delta_c=1.7$
(dashed curves) according to Mann \etal~(1993)'s prescription.

As noted by Mann \etal~(1993), the correlation functions computed using
$\delta_c=1.7$ consistently overestimate the correlation amplitudes on all 
scales of interest.  The correlation functions for $\delta_c=1$ are in 
excellent agreement with the numerical results for $D_c \leq 30h^{-1}$ Mpc 
(however, a value of $\delta_c \sim 1$ is rather unphysical).
For larger values of $D_c$, the analytic results tend to overestimate the 
correlations, with the discrepancy first becoming obvious on small scales 
and then propagating out to larger scales as $D_c$ continues to increase.   
For a given $D_c$, the discrepancy is more severe for SCDM07 clusters than for
O3CDM clusters.  

The tendency for the peak scheme to overestimate the correlations on
scales where $\xi_{CC}(r) \geq 1$ by a margin that grows larger with
increasing $D_c$ suggests that the peaks-based method will
overestimate the correlation lengths of samples with large values of
$D_c$.  Both Mann \etal~(1993) and Watanabe \etal~(1994) have computed
the $R_0$--$D_c$ relation predicted by the peaks method.  Comparing
their $R_0$--$D_c$ curves for O3CDM-like models with our numerical
results we find that for cluster samples with $D_c\approx 80h^{-1}$
Mpc, the correlation length predicted by the peaks-based
analytic scheme is $R_0\approx 28h^{-1}$ Mpc whereas the simulation
result is $R_0\approx 22h^{-1}$ Mpc, very close to the PS based
prediction.
The breakdown in the peak scheme may be due to several factors. The
simplest possibility is that that the method
uses the number density of clusters in a sample as a constraint rather
than some physical properties of the clusters such as their masses.
Other possible causes of the breakdown are: the
manner in which the different filtering scales are chosen, the
simplistic nature of the prescription defining the relation between
peaks in the smoothed density field and ``clusters'', and the
requirement that the large--separation asymptotic limit of the
statistical contribution to the cluster correlation function matches
the statistical peak--to--peak correlation function (Mann 1998,
private communications). The latter two tend to magnify any minor
discrepancy caused by any of the other factors.

\subsection{Correlation Length and the Cluster Abundance}

\begin{figure*}
\centerline{\hbox{
\psfig{figure=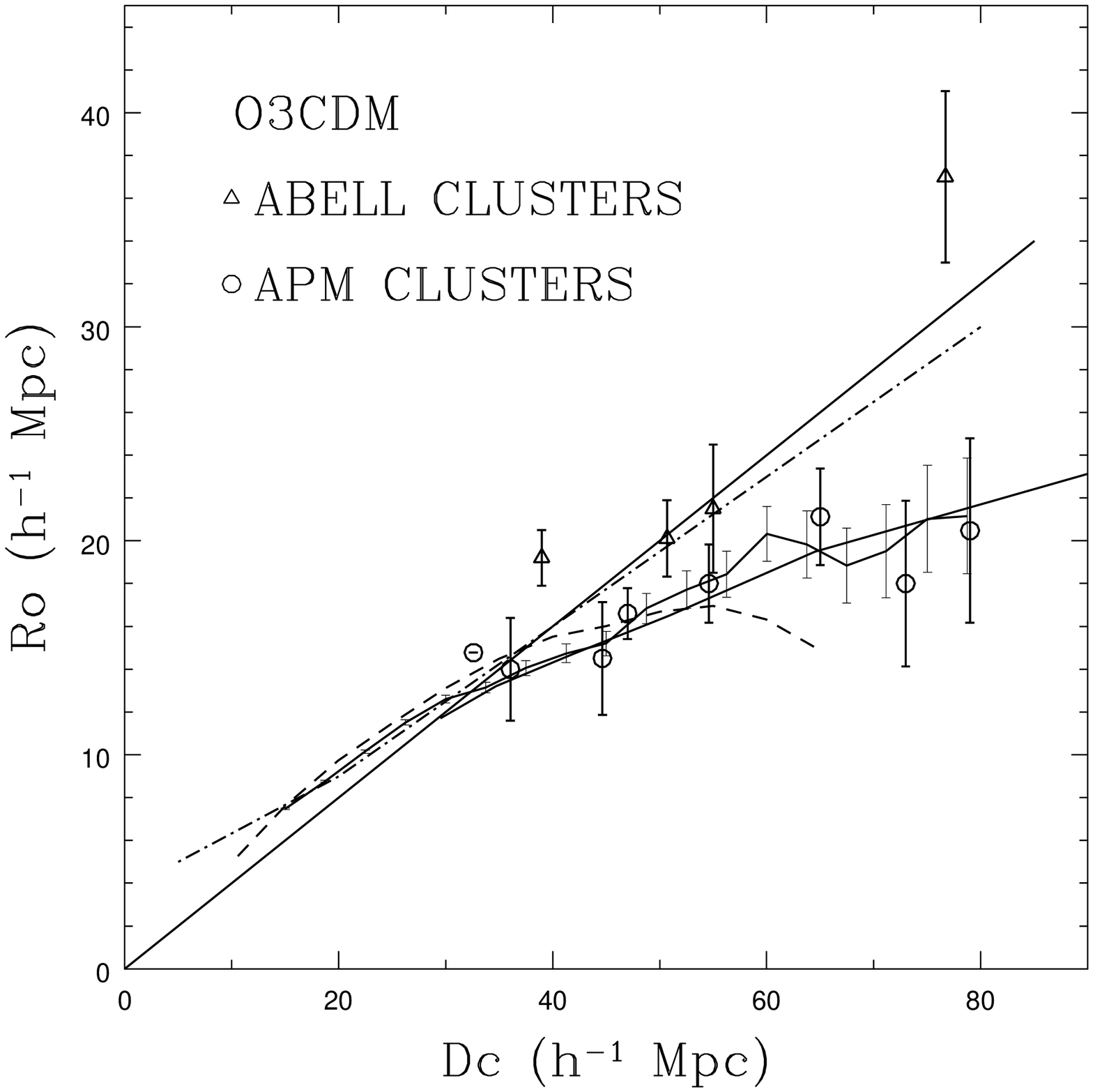,height=12cm,width=12cm,angle=0}
}}
\caption{Same as Fig. \ref{neta-scdm} but for O3CDM model.  
\label{neta-ocdm}
}
\end{figure*}

In Figures \ref{neta-scdm} and \ref{neta-ocdm}, we plot correlation
length ($R_0$) as a function of cluster abundance in terms of $D_c$,
the mean cluster separation, for the FOF clusters in the SCDM07 output and
O3CDM model at z$=$0. For comparison, we also show the scaling relation
(Equation \ref{bahcall-scaling}); the numerical results of Bahcall \&
Cen (1992) and Croft \& Efstathiou (1994) the observational data for
R$\geq 0$, R$\geq 1$, R$\geq 2$ Abell clusters (open triangles) from
Bahcall \& Soneira (1983) and Peacock \& West (1992) and the data for
the APM clusters (open circles) given by Dalton \etal~(1992) and Croft
\etal~(1997).  {\it In neither of the SCM07 nor  O3CDM models is the
$R_0$--$D_c$ relation for clusters consistent with the scaling
relation $R_0=0.4D_c$.}

However, it should be noted that the numerical results show the
$R_0$--$D_c$ relation for the real-space correlation function whereas
the correlation lengths for the observed clusters are derived from
redshift-space correlation functions.  Redshift--space correlation
lenghts are generally larger than their real-space correlation length
counterparts.  For example, in the case of their low-density spatially
flat CDM model, Croft \& Efstathiou found that their real-space
$R_0$-$D_c$ relation saturates at $R_0\approx 15h^{-1}$ Mpc for large values of
$D_c$, whereas the redshift-space $R_0$-$D_c$  saturates at $R_0
\approx 21 h^{-1}$ Mpc. An increase of this kind, however, is not
sufficient to bring our numerical $R_0$--$D_c$ relation into agreement
with the scaling relation of equation \ref{bahcall-scaling}.

Consistently with all previous findings, the $R_0$-$D_c$
curve for clusters extracted from the SCDM universe (Figure
\ref{neta-scdm}) does not match either the APM or Abell results.  On
the other hand, the results of our O3CDM model are in good
agreement with the APM and richness R$>0$, R$>1$ Abell cluster data,
even if the effect of redshift distorsions increasing the
length scale $R_0$  a few Mpc were included.

The seriously discrepant datapoint is  for R$>2$ Abell clusters.  If
this measurement is correct, it suggests that clustering on very large
scales may have been modulated by non-Gaussian processes (see Mann
\etal~1993; Croft \& Efstathiou 1994) as it is very difficult to
conceive of a Gaussian model that can produce the requisite clustering
at these  scales.  It would also imply that the very rich APM
clusters with their comparitively large correlation lengths are not
really rich or massive systems but rather are systems comparable to
R$>1$ Abell clusters whose number densities have been biased downward
by  the cluster identification algorithm.  
The agreement between our analytic results and
 our numerical results for  clusters in the O3CDM model and
the APM results leads us to believe that it is the R$>2$ Abell result
that is most likely incorrect, biased upward by the inhomogeneities
and contamination due to projection effects in the Abell catalog as
argued by Sutherland (1988) and Sutherland \& Efstathiou (1991).

In comparing our numerical results for SCDM07 to those of Bahcall \&
Cen (1992), we find that for $25h^{-1}$ Mpc $< D_c < 40h^{-1}$ Mpc,
our $R_0$-$D_c$ results are consistent with theirs. 
  For $D_c > 40h^{-1}$ Mpc, our curve rises less steeply than
that of Bahcall \& Cen and appears to saturate for $D_c > 50h^{-1}$
Mpc.  From analytic results (light solid curve), which we discuss
further in the next subsection, we expect the $R_0$-$D_c$ curve to
continue to rise but much more gently than the Bahcall-Cen result.
Since both we and Bahcall \& Cen (1992) used the FOF algorithm to
identify clusters in the simulations, the cluster selection algorithm
cannot be responsible for the differences.  Furthermore, our correlation
lengths were determined in the same way as Bahcall \& Cen (1992).

Comparing our O3CDM results to those of Bahcall \& Cen's (1992) low-$\Omega$
models, we find that the two are in good agreement for $D_c < 35h^{-1}$ Mpc
and also in a rough agreement with the scaling relation.  However as
the cluster abundance decreases and $D_c$ increases, the correlation lengths
of our cluster samples do not increase as quickly.  
Our numerical results  at large values of $D_c$ are in good  agreement
with  those derived analytically to the scales probed by our simulations.
(see \S3.3)

In comparing our $\sigma_8=0.7$ result to Croft \& Efstathiou's (1994)
$r_c=1.5 h^{-1}\,{\rm Mpc}$ $\sigma_8=0.59$ SCDM model, we once again
find smaller correlation lengths for $D_c < 30 h^{-1} \,{\rm Mpc}$.
 For higher values of $D_c$, the Croft \& Efstathiou (1994)'s results are
consistent with ours in spite of the fact that we have used FOF to
identify the clusters and Croft \& Efstathiou (1994) results are based
on a very different scheme.  Comparing our O3CDM results to those of
Croft \& Efstathiou's (1994) $1.5h^{-1}\,{\rm Mpc}$, $\sigma_8=1.0$
low-$\Omega$ spatially flat CDM model, we find that within the
uncertainties in the two curves, they are in excellent agreement with
each other.  The flattening  in  Croft \& Efstathiou's
curve for $D_c > 50 h^{-1} \,{\rm Mpc}$ (at $R_0\approx 15h^{-1}
\,{\rm Mpc}$) is not real.  As indicated by both our numerical and
analytic results, the correlation continues to rise, albeit gently,
reaching $R_0\approx 22h^{-1}$ Mpc at $D_c=80h^{-1}$ Mpc and is still
rising.  The flattening trend is  likely an artifact of the
finite simulation volume or even the poor mass/force resolution.

\section{THE CLUSTER MASS FUNCTION}

According to the analytic PS formalism, the comoving number density of 
dark matter halos of mass $M$ in the interval $dM$ is
\begin{equation}\label{diff-ps-mass}
{N(M)}=\sqrt{2\over\pi}{\bar\rho\over M^2}
{\delta_cD^{-1}\over\sigma}\left\vert{d\ln\sigma\over d\ln M}\right\vert
\exp\left[-{\delta_c^2D^{-2}\over 2\sigma^2}\right],
\end{equation}
where $\bar\rho$ is the comoving density of the Universe and $\sigma(M)$
is the linearly extrapolated present-day $rms$ density fluctuation in spheres
containing a mean mass $M$.  The redshift evolution of $N(M)$ is controlled 
by the density threshold for collapse, $\delta_c/D(z)$, where $D(z)$ is 
the linear growth factor normalized to unity at $z=0$ (Peebles 1993) and 
$\delta_c$ is the linearly evolved density contrast of fluctuations that are
virializing at $z=0$. The growth factor, $D(z)$, depends 
on $\Omega_0$ and $\Lambda$ whereas $\delta_c$ has only a weak 
dependence on $\Omega_0$. For spherical density fluctuations,
$\delta_c = 1.686$ for $\Omega_0=1$ and $1.65$ for $\Omega_0=0.3$.

The PS description of structure formation in the Universe leads 
naturally to the definition of a characteristic mass $M_*(z)$ such that 
\begin{equation}\label{charmass}
\sigma(M_*)D(z)=\delta_c\, .
\label{m*}
\end{equation}
$M_*(z)$ is then the characteristic mass of halos that are virializing
at redshift $z$.  Its evolution tracks the manner in which structure
forms.  In  bottom-up hierarchical clustering models, such as CDM
models, $M_*(z)$ increases as a function of time as lower mass
structures are incorporated into progressively more massive halos.  In
a critical universe, the growth factor evolves as $D(z)=(1+z)^{-1}$
and to first order, this implies a strong evolution in $M_*$.  In an
open or a flat, low-$\Omega_0$ universe, $D(z)$ ceases to evolve as
strongly, and the evolution of the characteristic mass is greatly
suppressed, once $\Omega(z)$ deviates significantly from unity.
Hence, the evolution of the dark halo mass distribution is also
greatly suppressed.  A clear detection of the presence or absence of
strong dynamical evolution in the cluster population can be used to
put stringent limits on the underlying cosmology.

The actual value and the details of the evolution of $M_*$, and
therefore of the mass distribution especially for $M > M_*$, depends
sensitively on $\delta_c$.  The standard practice is to use the value
of $\delta_c$ for collapse of spherical perturbations. Typical
perturbations in CDM models, however, are not spherical and therefore,
the actual value of $\delta_c$ will differ from the spherical value.
Moreover, Heavens \& Peacock (1986), argue that $\delta_c$ is likely
to be lower ($1 \leq \delta_c \leq 1.68$) because typical
proto-structures in a Gaussian random field tend to be triaxial.  {\it
A lower (higher) value of $\delta_c$ results in more (fewer) high mass
objects}.  A detailed discussion of $\delta_c$ and the asphericity of
the density perturbations is given by Monaco (1995, 1998), who finds
that when the assumption of spherical collapse is relaxed, $\delta_c$
 becomes a function of the local shape of the perturbation
spectrum.  In most cosmological models, the power spectrum of the
primordial perturbations over the scales of interest deviates, albeit
gently, from a simple power law shape and it becomes debatable whether
a constant value of $\delta_c$ is a fair description of the evolving
cosmic mass function at all masses and redshifts.  Here we allow the
collapse threshold to be a free parameter depending on redshift,
calibrating at each epoch using the high mass end of the mass
distribution.  As discussed in \S1 and \S2.3, the halo mass function
derived using the PS formalism is a measure of the abundance of
collapsed, distinct, halos characterized by their virial radius and
mass.  Consequently, it is appropriate to use catalogs generated
using the FOF and HOP cluster finding algorithms.

\subsection{Computing the halo mass function}

We construct the differential mass function by sorting the halos
according to their masses in bins of size $\Delta$log(M) = 0.1.  We
have verified that our results are insensitive to this choice of bin
size.  Due to the large size of our simulations, we are able to study also
the {\it differential} cluster mass function instead of the cumulative
distribution, as is usually done.  This means that the individual bins
are independent and the results more robust.  We estimate the
uncertainty in the number of objects in each bin using Poisson
statistics. It is useful to remember that the SCDM run can be
rescaled freely to a different $\sigma_8$ normalization, The redshift
of each given output is then rescaled to a redshift z':
$ 1+z' = (1+z)/\sigma_8$.

\subsection{ Effects of numerical resolution and cosmic variance}

To study the effects of degrading the numerical resolution, we
examined a lower resolution run of our SCDM07 volume. This run used
the same phases, 3 million particles, a softening length of
$160h^{-1}$ kpc and a third of the timesteps used in our fiducial run.
Both force and spatial resolution are therefore significantly poorer.
Halos represented by 128 particles in our fiducial run have only 8
particles in the low resolution simulation. While there is good
agreement at the high mass end, the low mass end of the mass function
is severely affected by the poorer resolution, showing a significant
decrease in the number density of halos below $ 3 \times 10^{14}
h^{-1} M_\odot$.  This test indicates that at least 30 particles are
needed to correctly assign a mass to individual halos. Our choice to
include in our analyses only halos with N$>64$ is then a conservative one.

Finally, we also explored the effects of cosmic variance on the
cluster number density.  We divided our $z=0$ SCDM07 volume into
several subvolumes and measured the local $\delta_c$ over the same
mass range as we did for the whole volume.  As expected, cosmic
variance produces a scatter in $\delta_c$ when measured in smaller
volumes.  However the scatter is not significant for volumes as little
as $1/8$-th of the original simulation volume (i.e.~cubes with
250h$^{-1}$Mpc per side): we find $\delta_c$ = 1.68 $\pm$ 0.02. This
is close to the error associated with a single measurement and similar
to the value we get from the whole volume, suggesting that the value
of $\delta_c$ for a given halo finder has (almost) converged when the
full volume of the simulation is considered.

\subsection{Press-Schechter Predictions vs Numerical Results}

\begin{figure*}
\centerline{\hbox{
\psfig{figure=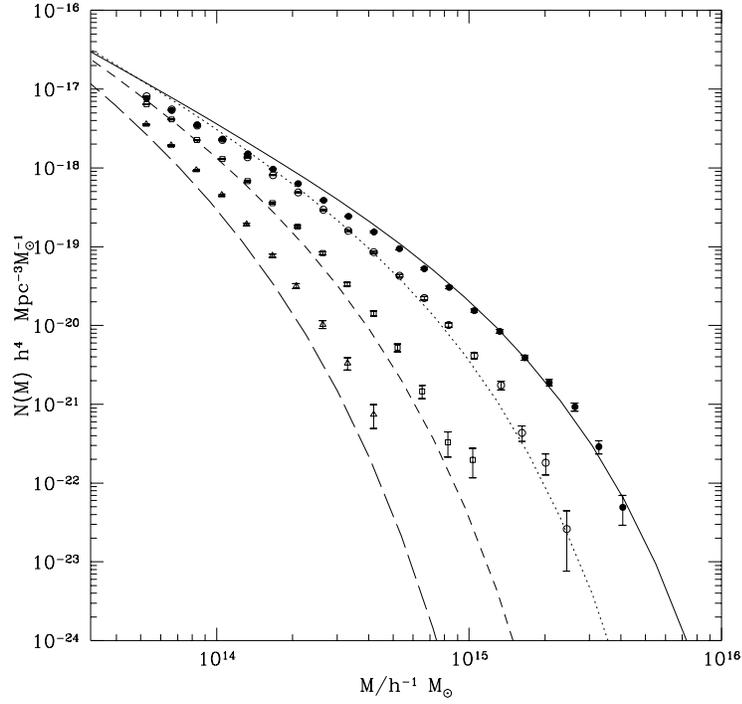,height=10cm,width=10cm,angle=0}
}}
\caption{ The N-body mass distribution for the SCDM run --- shown as
points --- at four different outputs as well as the standard PS mass
function (computed using $\delta_c=1.686$) --- shown as lines --- for
the same four outputs: $\sigma_8=1$ (filled circles, solid line),
$\sigma_8=0.7$ (circles, dotted line),$\sigma_8=0.47$ (squares, dashed
line), $\sigma_8=0.35$ ( triangles, long dashed line).  The error bars
correspond to 1$\sigma$ Poisson errors.  
\label{scdm_mf}
}
\end{figure*}

\begin{figure*}
\centerline{\hbox{
\psfig{figure=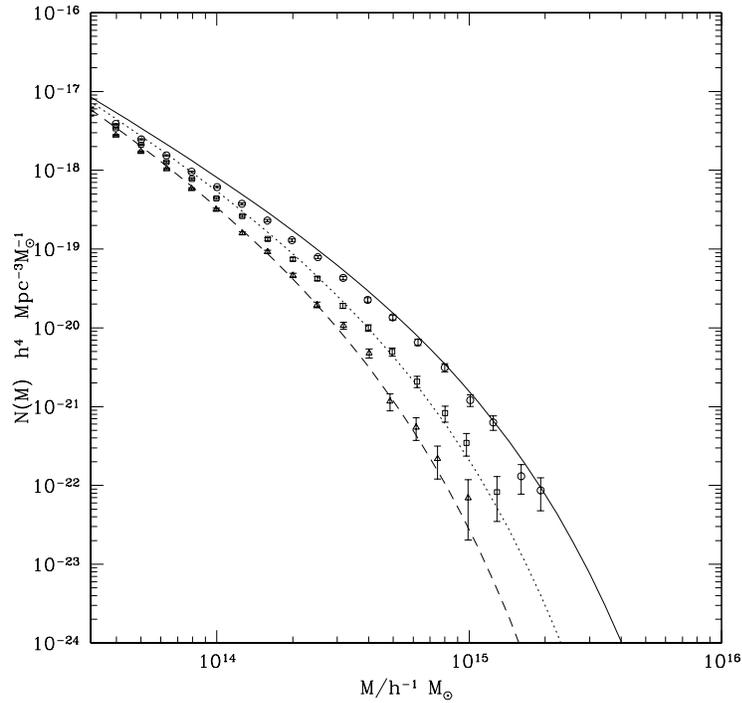,height=10cm,width=10cm,angle=0}
}}
\caption{
The same as Figure \ref{scdm_mf} but for O3CDM model at z=0,0.58,1. 
\label{ocdm_mf}
}
\end{figure*}

In Figures \ref{scdm_mf} and \ref{ocdm_mf}, we show the differential
mass functions from the SCDM and O3CDM simulations, respectively.  The
corresponding PS curves, computed using the canonical value of
$\delta_c$ for spherical perturbations, are also shown.  We only show
the FOF results.  Generally, the HOP and FOF results are very similar,
with HOP having a slightly larger number of massive clusters.

In the case of the SCDM model, we find that the shape of the
differential PS mass function is roughly consistent with the shape of
the numerical mass function only at $\sigma_8=1$ At lower $\sigma_8$
(or alternatively, higher z) the PS mass function {\sl underestimates}
the number density of {\sl rich clusters} in the simulation. The
excess at the high mass end is about a factor of a few in number
density per mass bin.  At $\sigma_8$ 0.7 or larger the PS approach
overestimates the number of small halos (M $< 10^{14}h^{-1}\;
M_\odot$). The deficit of low mass halos (which we will only touch
upon briefly here) has been well-documented in numerical works by
Carlberg \& Couchman (1988) and more recently, by Lacey \& Cole
(1994), Gross \etal~(1998) and Somerville \etal~(1998).  This deficit
arises independently of the choice of algorithm used to define the
halos in the simulations (see Figure \ref{hfcomp}) and has been
associated with merger events not accounted for by the PS formalism
(see Cavaliere \& Menci~1997 and Monaco~1997).  The fact that the two
halo finders agree extremely well in this regime makes the result very
robust.

Apart from the above-mentioned deficit, most other studies that have
tested the analytic PS mass function against numerical results have
reported a good agreement between the two (e.g. Eke \etal~1996) with
the notable exceptions of Bertschinger \& Jain (1994) and Sommerville
\etal~(1998), who found a systematic excess of massive halos in
numerical simulations as compared to the PS prediction.  These
numerical results, however, are derived from simulations that have
lower resolution and probe smaller cosmological volumes than our
simulations.  These simulations, consequently, contain only a small
number of massive clusters and this, in conjunction with cosmic
variance, has resulted in numerical mass functions with large
uncertainties at the high mass end.  Eke \etal~(1996), found a good
agreement with the PS mass function on the scale of $5\times
10^{14}h^{-1} M_\odot$ but with a large error bar.

The large volumes used in our work allow us to confirm the good
agreement between the numerical and the analytical mass function found for
high values of $\sigma_8$.  The excess of massive clusters at lower
$\sigma_8$/higher redshifts suggest that the cluster mass function for
the SCDM model evolves more slowly than the PS mass function.  At
$\sigma_8$=0.35, the number density of Coma--like clusters exceeds the
standard PS prediction by almost an order of magnitude.  Allowing
$\delta_c$ to be a free parameter at each epoch, we perform a $\chi^2$
fit of the PS formula to the numerical results assuming Poisson errors
for each bin.  Since observations tend to be biased towards the
high-mass end, we include in our fit only clusters with temperature
T$>$3keV, (the M--T relation is defined in \S4.4) to allow the use of
this formula in theoretical predictions for X-ray clusters. This
formula will underestimate slighlty the number of very (T$>$7keV) hot
clusters for high $\sigma_8$ models.

\begin{figure}
{\epsfxsize=8.2truecm \epsfysize=8.2truecm 
\epsfbox[20 150 580 730]{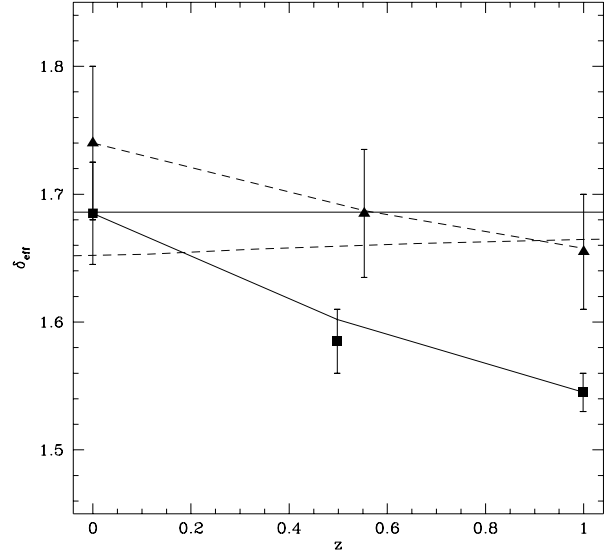}}
\caption{ The points show best fit values of $\delta_{c(z)}$ required
to match the numerical mass function for clusters with T$>3$ keV at
different redshifts.  The squares show the results for the SCDM07
model and the triangles show the results for O3CDM model.  The
horizontal light solid curve and the nearly horizontal dashed lines
show the canonical values of $\delta_c(z)$ for the SCDM07 and the
O3CDM models respectively.  The lines across the points are the
power-law interpolation to the points (see text).  Bars are 3 $\sigma$
errors.
\label{dc}
}
\end{figure}

For the SCDM cluster mass function at $\sigma_8$=0.7  , the best--fit
$\delta_c$ coincides with the canonical value of $1.686$~!
For $z>0$ and for our FOF selected halos, the best-fit value of
$\delta_c$ as a function of redshift and $\sigma_8$ is :

\begin{equation}
\delta_{c}(z) =  
1.686
\left[
\left(\frac{0.7}{\sigma_{8}}\right) 
(1+z)
\right]^{-0.125} 
\label{eq:dc}
\end{equation}

as shown in Figure \ref{dc}, whereas the canonical value of $\delta_c$
is a constant. (in Figure \ref{dc} errorbars are 3 $\sigma$ errors).
For HOP selected halos $\delta_c$ is offset toward even lower values,
i.e. toward a larger cluster excess: $\delta_c = 1.6$ at z$=$0. (for
the SCDM model at $\sigma_8 =0.7$).  However, a similar evolution of
$\delta_c(z)$ is found, with a $ (1+z)^{-0.1}$ dependence.  In
principle the fitting procedure should keep into account that the mass
associated with a given temperature depends on z and that the same
output can be associated with different z depending on
$\sigma_8$. However the fitting formula is not affected by this for
values of present day $\sigma_8$  between 1 and 0.5 and  eq. \ref{eq:dc} can
be used safely.

In the O3CDM model case , the PS mass function is in fair
agreement with the numerical mass function, especially at low redshifts. 
At $z=0$, the high mass end of the numerical halo mass function agrees 
within 2$\sigma$ with the analytical curve computed using the 
canonical (spherical) value of $\delta_c$ = 1.651.  The uncertainties 
in the number densities of massive clusters are slightly larger in the 
O3CDM case because of the smaller physical volume/h igher H$_0$ of the 
simulation.  As shown in Figure \ref{dc}, the best-fit $\delta_c(z)$ for FOF 
halos can be well-approximated as

\begin{equation}
\delta_{c}(z) =  
1.775 
(1+z)
^{-0.07} 
\label{eq:dc2}
\end{equation}

HOP results shows an even weaker evolution in z.  As the result is
more significant in the SCDM case we will mainly focus on the analysis
of results for the critical case.

\begin{figure}
{\epsfxsize=8.2truecm \epsfysize=8.2truecm 
\epsfbox[20 150 580 730]{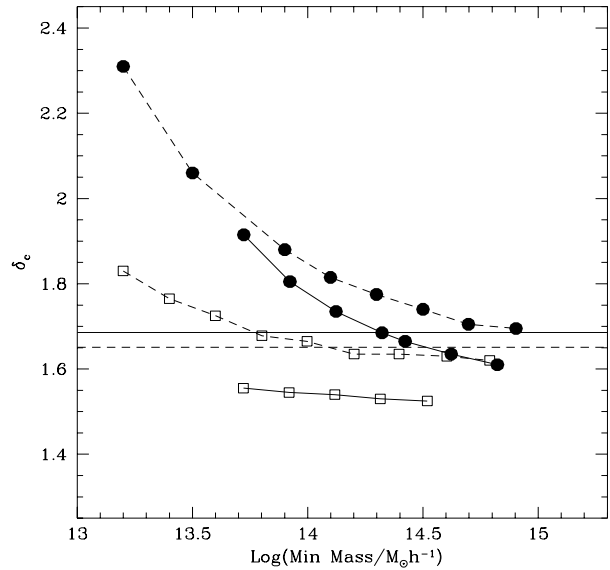}}
\caption{The points show the best-fit values $\delta_c(z)$ for SCDM07
(continuous) and O3CDM (dashed) at z=0 (black dots) and z=1 (i.e
$\sigma_8=0.35$ for SCDM) (open squares) as the mass range over which the PS
functional form is fitted to the numerical FOF mass function is
varied.  The abscissa corresponds to the lower mass threshold of the
mass range over which the fit is demanded.
\label{DeltaCm}
}
\end{figure}

From these results however, it is not clear if the deviation from the
standard PS mass function is due to just one or rather both of the
following effects:

$\bullet$ at lower $\sigma_8$ and for a fixed temperature T we study
more extreme clusters i.e. we look at a different region of the mass
function, which maybe still be self similar, albeit different from the
canonical PS.

$\bullet$ the shape of the mass function evolves with time
 and/or depends on the  power spectrum

As discussed above, the best-fit $\delta_c(z)$ were determined by
fitting the functional form of the PS mass function to the numerical
results for clusters with T$>3$ keV, i.e.  on mass scales $M \gg M_*$,
where $M_*$ corresponds to $4\times 10^{13}\, h^{-1} M_\odot$ for
SCDM07 and $1.5\times 10^{13}\, h^{-1} M_\odot$ for O3CDM.  It is of
interest to relax this constraint and explore how the $\delta_c$
varies as the minimum mass of the halos included in the fit is lowered
and approaches the mass of the smallest halos (64 particles) in our
catalog. We carried out the above exercise at two different
epochs/normalization and the results are shown in Figure
\ref{DeltaCm}.  The value of $\delta_c$ changes dramatically as the
mass range over which the fit is carried out moves toward smaller
masses (the fit is dominated by the smaller mass bins as they contain
most of the halos used in the fitting).  The trend for both SCDM and
O3CDM is for $\delta_c$ to become larger as the mass threshold is
lowered.  This is precisely what one expects given the deficit of
low-mass halos in the simulations.  This shows that the shape of the
N--body mass function differs from the PS prediction, and the exact value of 
$\delta_c$ is a function of the mass interval considered.

\begin{figure*}
\centerline{\hbox{
\psfig{figure=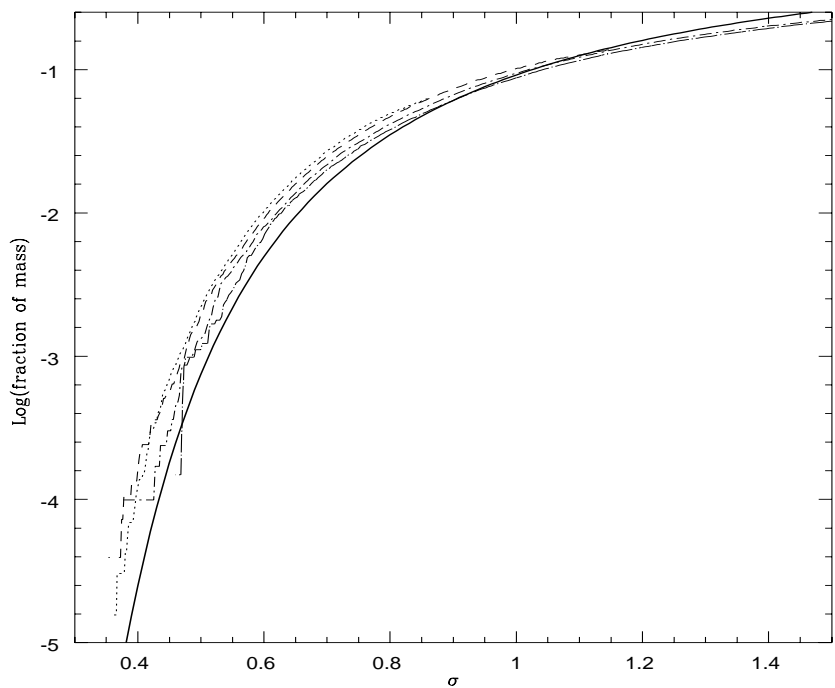,height=10cm,width=12cm,angle=0}
}}
\caption{The plot shows the cumulative fraction of mass in collapsed
halos as a function of the variance of the density field.  The PS
prediction (invariant with redshift and cosmology) is the thick
continuous line. The SCDM results are shown at diffferent outputs:
$\sigma_8=1$: dot--long dashed; $\sigma_8=0.7$: dot--short
dashed; $\sigma_8=0.47$: dottted line; $\sigma_8=0.35$: dashed. The
numerical mass function is obviously not invariant and shows evidence
of evolution.
\label{pscumul}
}
\end{figure*}

This plot shows also that, at a given mass, $\delta_c$ is a function of
redshift. To prove that the shape of the mass function evolves with
time we take advantage of the fact that within the PS framework the
cumulative fraction of mass in collapsed halos is invariant when
plotted vs the variance of the density field at a given mass/lenght
scale. I.e. at a given value of $\sigma$ (which corresponds to
different mass/lenght scales depending on cosmology and z) the
fraction of mass in collapsed objects is always the same. This  is
shown in Figure \ref{pscumul}.

It is interesting to interpret this change in terms of different power
spectra. SCDM models can be rescaled to models with different
$\Gamma'$ (or $\tau$CDM models) rescaling the box by $\Gamma/\Gamma'$
and choosing as the final output the one with the correct present--day
normalization at the scale  corresponding to the 8h$^{-1}$Mpc
scale. The sequence of outputs with decreasing $\sigma_8$ and increasing
excess of massive objects can be reinterpreted as a sequence of models
with smaller $\Gamma$ parameter and the same normalization, revealing
the dependence of the shape of the mass function from the power
spectrum. An excess of massive halos for a more negative local spectral
index (as the case with larger $\Gamma$) had been predicted by Monaco
(1995).

This rescaling allowed to compare our mass functions with that
obtained from the so called ``Hubble Volume Simulation'' (HVS)
(Colberg \etal~1998), a $\tau$CDM with $\Gamma$=0.21 and normalization
of $\sigma_8=0.6$ Their final output can be rescaled to a SCDM model
with a $\sigma_8 =$ 0.285. The HVS mass function lies nicely along our
sequence of mass functions (Cole \& Jenkins 1998, private
communication) showing a slightly larger excess of halos compared to
our $\sigma_8 = 0.35$ output (our output with the lowest $\sigma_8$).

 The two simulations used completely independent software to generate
the initial conditions, evolve the density field and analyze the data.
The agreement found is quite satysfying and shows that the results of
both simulations are free of hidden systematic effects.

{\it We can conclude that for critical CDM models the numerical mass
function compared to the PS analytical prediction has an excess of
halos for M$\gg M_*$ and a deficiency for masses M $\ll M_*$. The
excess at large masses is larger for models with smaller $\Gamma$ (or
more negative local spectral index)}.

These deviations for the canonical predictions are significant and
cosmological tests based on the number density of a particular class
of objects need to use the appropriate value of $\delta_c$ to make
robust predictions. Our fitting formula eq. \ref{eq:dc} can easily be
modified for other critical models with a different shape parameter.

\subsection{Effects on the  Cluster Temperature distribution} 

X-ray observations allow one to directly determine the cluster
temperature or the cluster luminosity function. Of the two, the
temperature of the intracluster gas is thought to be the more robust
measure of the depth of the cluster potential well and to a  good
approximation, is expected to be very strongly correlated with the
cluster mass.  In this section, we examine the impact on the cluster
temperature function of the excess of massive clusters found in the
simulation as compared with the predictions of the standard PS mass
function.

We follow Eke, Cole \& Frenk (1996) and adopt 
the following simple relation for estimating the temperature of the 
intracluster medium of a cluster of  mass M :
\begin{equation}
k T = 7.75 M_{15}^{2/3}(1+z)\Big({{\Omega_o \Delta_c}\over {\Omega(z) 178}}
\Big)^{1/3}.
\label{mt}
\end{equation}
$\Delta_c$ is the average density contrast at virialization with respect 
to the critical background density.  This relation assumes that the
intracluster medium is isothermal. M$_{15}$ is  mass in units 
of 10$^{15}M_\odot h^{-1}$

\begin{figure*}
\centerline{\hbox{
\psfig{figure=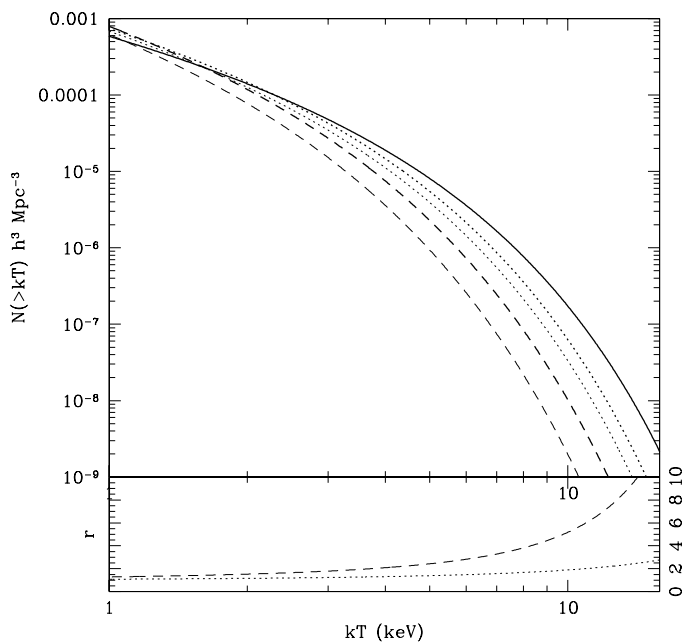,height=10cm,width=10cm,angle=0}
}}
\caption {Upper panel: PS (thick lines) vs numerical (thin lines) cumulative
 temperature function N($>$T) for the SCDM07 model at $z=0$ (continuous
 line), $z=0.5$ (dotted), and $z=1.0$ (dashed). Bottom panel:
 shows the ratio of clusters above a given temperature between the
 numerically-based temperature function and the PS-based temperature
 function at $z=0.5$ (dotted), and $z=1.0$ (dashed). 
 }
\label{scdmnt}
\end{figure*}

The cumulative temperature function $N(>T)$ considering the SCDM07
model as the present time is shown in Figure \ref{scdmnt}.  The
corresponding cumulative temperature function based on the standard PS
mass function is shown by the dashed lines. The bottom panel shows the
ratio between these two mass functions.  At $z=0$, the two are very
similar.  At $z\sim 1$, however, the number density of clusters with
temperatures $kT > 7$ keV (the temperature of rich, Coma-like
clusters) is more than a factor of 3 greater than the canonical PS
prediction and in the case of exceptionally hot clusters ($kT>10$
keV), the discrepancy is an order of magnitude. This discrepancy 
increases for a universe with a lower, more realistic normalization.

One implication of this is that the cumulative temperature function
obtained from the simulation evolves more slowly over the redshift
range $0<z<1$ than PS theory predicts. Since we choose a fixed
temperature range, at higher z we are observing more extreme (massive)
clusters, i.e. a region of the mass function with a stronger excess
compared with the anlytical formula.  In principle it may be more
difficult to challenge a high $\Omega$ model on the basis that the
observed cluster temperature function does not vary significantly over
the redshift range $0 < z < 0.3$, especially if this effect is coupled
with a significant scatter in the M--T relation.  The higher number
density of hot clusters in the simulations also suggests that attempts
to estimate $\sigma_8$ by fitting the canonical PS-based temperature
function may in principle lead to systematically high results.

Following Eke {\it \etal} (1996, 1998) we reanalysed the low redshift
cluster data of Henry \& Arnaud (1991) and the new data in Henry {\it
\etal (1997)} to produce a cumulative cluster temperature function. We
then compared it with PS prediction modified using our inferred values for 
$\delta_c(z)$.
This leads to a small revision of the amplitude of fluctuations inferred
from the local cluster abundance for $\Omega_{0}=1$, to a value of
$\sigma_{8} = 0.5 \pm 0.04$ if FOF halos are used.   The use of HOP halos
 results  suggests $\sigma_{8} = 0.48 \pm 0.04$. 

{\it With a $\sigma_{8}$ = 0.5 normalization and including the
 rescaling of $\delta_c$ (eq \ref{eq:dc}) the excess of hot clusters
 with $kT > 7$ keV compared to the standard PS prediction amounts
 almost to a factor of 10 at redshift of 1}.  Clearly, claims that
 rule out critical density models on the basis of the detection of a
 single massive cluster at high z (Donahue {\it \etal} 1998) must then
 be taken with caution.  However, the discrepancy between the
 temperature function predicted in a critical density universe and
 that observed at $z=0.33$ is reduced by a modest amount using the
 modified Press-Schechter scheme.  The discrepancy there is still
 large enough to rule out $\Omega_{0} = 1$, unless there are
 significant differences in the relation between mass and temperature
 for clusters at high and low redshift that make equation \ref{mt}
 invalid.

\section{Conclusions and discussion}

We have analyzed parallel N--body simulations of three CDM models: a 
critical density model (h=0.5) and two open models
($\Omega_0$ =0.3 and 0.4).  These three  models,
span a large range of different properties (models COBE and cluster
normalized, different amounts of large scale structure, high and low
$H_0$) to cover the wide range of cosmological models presently
considered.  We simulated very large volumes -- 500h$^{-1}$Mpc per side --
and used 47 million particles for each run. Having good mass, force
and spatial resolution, these datasets allow a robust determination
of two very important quantities for cosmological studies: the
correlation function of galaxy clusters and the shape and evolution
of the cosmic mass function on  cluster mass scales. In particular,
we have focused on how results from simulations compare with predictions
from analytical methods based on the PS approach.

$\bullet$ {\it Halo finders} We have assessed if cluster correlations
and mass functions are affected by the biases introduced by using
different halo finders.  This is an important step both for comparing
the simulation results with theoretical predictions as well as with
observational data.  We used two halo finders available in the public
domain: FOF and HOP.  We found that the two halo finders, once set to
select virialized structures based on a local overdensity criterion,
produced very similar halo catalogs. While a small mass offset is
detectable, (of the order of 5--7\% for massive clusters) our
conclusions do not depend on the particular choice of the halo
finder. In most of our work we conservatively used FOF, which gives
results closer to the analytical approach.

$\bullet$ {\it Cluster correlation function} We were able to determine
the clustering properties of halos spanning nearly two orders of
magnitude in mass, ranging from groups and poor clusters to very rich
massive clusters.  Our analysis has shown that the {\it comoving}
correlation functions, for a wide range of group/cluster masses, do
not change significantly over the redshift range $0 < z < 0.5$.  Of
the two analytic schemes we used (peaks-based and PS-based) for
computing cluster correlations functions, we found that
the correlation functions derived using the PS-based scheme (Mo, Jing
\& White 1996, Baugh \etal~1998) were in excellent agreement with the
numerical correlation functions.

$\bullet$ {\it Cluster Correlation Length and Number Density} We
compared our results for the cluster $R_0$--$D_c$ relation in
different models against both observations and results of previous
numerical studies.  Firstly, we have confirmed that the SCDM model
does not have sufficient large scale power to account for the observed
clustering of clusters.  The $R_0$--$D_c$ relation for clusters in the
low density models, is consistent with the scaling relation
$R_0=0.4D_c$ for $D_c < 50h^{-1}$ Mpc.  On larger scales the cluster
correlation length increases more slowly than the cluster number
density.  We also found that generally our $R_0$--$D_c$ results were
consistent with those of Croft \& Efstathiou (1994) in spite of the
fact that we used FOF to identify the clusters in the simulations and
Croft \& Efstathiou (1994) used a very different algorithm.  The one
difference between our results and those of Croft \& Efstathiou (1994)
is that in their low $\Omega_0$ run that $R_0$ becomes constant for
large values of $D_c$ whereas we found that $R_0$ continues to
increase, albeit gently.  The analytic $R_0$--$D_c$ relation based on
the PS approach agrees with our numerical results. The flattening of
the $R_0$--$D_c$ is qualitatively expected in CDM models, where the
bias of halos depends rather weakly on mass (a factor of $\sim 2$ when
the mass changes by a factor of 100, see Table \ref{table:bias2},
while at the same time the number density of massive clusters
decreases exponentially fast. This implies that, as clusters of
increasing mass ( i.e. larger $D_c$) are considered, $R_0$ (which is
linked to the bias) grows slower than requested by the Bachall \& West
relation.

Finally, we found that the $R_0$--$D_c$ relation for our low-density
CDM model is consistent with the results for richness R$>0$ and R$>1$
Abell clusters as well as those for the APM clusters, including those
for the very rich APM clusters.  The only disagreement is with the
single datapoint for R$>2$ Abell clusters; the correlation length for
this sample is too large compared with the APM measurements and the
N-body results.  This would  suggest some strong systematic
differences (or selection criteria) between rich APM clusters and rich
Abell clusters with the same number density.  As discussed above, it
does not seem possible to reconcile CDM models with the Bachall \&
West relation, unless some exotic process has boosted the bias of
massive clusters compared to the dark matter distribution. These
considerations hold also for flat models with a cosmological constant,
as suggested by preliminary results by Colberg \etal (1998).

$\bullet$ {\it Cluster mass function} The analytical PS prediction
differs from the N-body mass function both at the low and high mass
ends (i.e. for both M/M$_*<<1$ and {M/M$_*>>$1}).  These discrepancies
are more significant in the critical SCDM universe. Consequently, the
analytic cumulative cluster temperature function based on the standard
PS mass function underestimates the abundance of hot clusters.  The
discrepancy gets larger with lower $\sigma_8$ and/or lower values of
$\Gamma$ ( large scale structure studies suggest $\Gamma \sim 0.25$
(Maddox \etal~1990)). At $z=1$ (assuming a SCDM cluster normalized
model with $\sigma_8$ = 0.5) the number of clusters with kT$>7$ keV is
underestimated by almost a factor of ten.  Claims that rule out
critical density models on the basis of the detection of massive
clusters at high z (Donahue \etal 1998) should then be taken with some
caution.  The temperature function obtained from our simulations
evolves more slowly than the standard PS temperature function.  These
results however, are not sufficient to bring the cluster temperature
function for $\Omega_{0}=1$, CDM models in agreement with that
observed at $z=0.33$ by Henry {\it \etal} 1998).

Our results show  there is no evidence to support the
notion of a universal value for $\delta_c$ in a given cosmology for
different power spectra.  Our results strongly support the idea that
$\delta_c(z)$ is a function of cosmology, and also of redshift and of the
mass range under consideration. It is then harder to attach a simple
physical meaning to the value of $\delta_c(z)$.  We give a fitting
formula for $\delta_c(z)$, valid for groups and clusters that improves
the agreement between the PS and numerical mass functions. It is
tempting to interpret the excess of high mass objects found in the
SCDM simulation as a deviation of the gravitational collapse from
the idealized spherically homogeneous collapse model (Monaco 1995).

One important, general consideration that can be extracted from our
  analysis is the primary importance of comparing results on
  homogeneous grounds.  In particular, it is crucial to define cluster
  samples on the basis of the same, physically motivated quantity. If
  clusters are defined in terms of the mass within the virial radius,
  then both analytical and numerical results can be directly compared
  and the resulting differences understood.  This makes theoretical
  predictions much more robust and easier to compare with
  observational results.
 
It appears that collisionless large scale structure simulations are
now in their maturity. Simulations even larger than ours are currently
being analyzed (see Colberg \etal 1998).  The statistical results
appear to be robust and stable, allowing a fruitful comparison with
observational data, which at the same time, are about to experience a
manifold growth. We are optimistic that a close interplay between
theoretical and observational results about the internal physics of
galaxy clusters and their large scale distribution will help to
improve our knowledge of the physical state of our Universe.

\section*{Acknowledgments}
We thank Shaun Cole and Adrian Jenkins for helpful discussions that
greatly improved our understanding of the results  in \S4.  We are
grateful to Joachim Stadel, one of the fathers of PKDGRAV. We thank
A.Cavaliere and P.Monaco for many useful discussions.  We thank Bob
Mann for his help in computing analytic peaks-based cluster
correlation functions.  We thank Rupert Croft and Renyue Cen from
providing us with their numerical results from their respective
studies. F.G acknowledges support from the TMR Network for the
Formation and Evolution of Galaxies. A.B. would like to thank the
Department of Astronomy at University of Washington for hospitality
shown to him during the summer of 1998 and gratefully acknowledges
financial support from University of Victoria and through an operating
grant from NSERC.  P.T. acknowledges hospitality from the Astrophysics
Department of the University of Durham and support from a NASA
ATP--NAG5--4236.  CMB and FG would like to thank the organisers of the
Guillermo Haro workshop on the Formation and Evolution of Galaxies,
held at INAOE, Mexico for their hospitality whilst this paper was
finished.

\label{lastpage}

\end{document}